\documentclass[12pt,epsfig,epsf]{article}
\usepackage{graphicx}
\usepackage{amssymb,amsbsy}
\usepackage{amsmath}
\usepackage{epsfig}
\usepackage{caption}

\def\be{\begin{equation}}
\def\ee{\end{equation}}
\def\bfi{\begin{figure}}
\def\efi{\end{figure}}
\def\bea{\begin{eqnarray}}
\def\eea{\end{eqnarray}}

\begin{document}

\begin{center}
{\Large \bf Energy and Heat Fluctuations in a Temperature Quench} \\
\vskip1cm
M.Zannetti$^{1,2}$, F.Corberi$^1$, G.Gonnella$^3$ and A.Piscitelli$^4$\\
\vskip0.5cm
$^1${\it Dipartimento di Fisica ``E.Caianiello'' and CNISM, Unit\`a di Salerno,\\
Universit\`a di Salerno, via Giovanni Paolo II 132, \\
84084 Fisciano (SA), Italy} \\
$^2${\it Kavli Institute for Theoretical Physics China, CAS, Beijing 100190, China}
$^3${\it Dipartimento di Fisica, Universit\`a di Bari and INFN, \\
Sezione di Bari, via Amendola 173, 70126 Bari, Italy} \\
$^4${\it Division of Physical Sciences, \\
School of Physical and Mathematical Sciences, \\
Nanyang Technological University, 21 Nanyang Link, 637371, Singapore.} 
\end{center}

\begin{abstract}

Fluctuations of energy and heat are investigated during the relaxation following
the instantaneous temperature quench of an extended system. Results are obtained analytically
for the Gaussian model and for the
large $N$ model quenched below the critical temperature $T_C$. The main finding is that fluctuations exceeding
a critical threshold do condense. 
Though driven by a mechanism similar to that of Bose-Einstein condensation,
this phenomenon is an out-of-equilibrium feature produced by the
breaking of energy equipartition occurring in the transient regime.
The dynamical nature of the transition is illustrated by phase diagrams extending
in the time direction.

\end{abstract}

\vskip2cm

\tableofcontents

\section{Introduction}

Non equilibrium statistical mechanics is a vast subject, rapidly evolving under the
push of major recent breakthroughs, like the discoveries of fluctuations theorems, the 
out-of-equilibrium generalizations of the fluctuation-dissipation relations
or the development of variational principles~\cite{Seifert}. In this paper
we focus on the fluctuations of macrovariables in non-equilibrium conditions,
analyzing the remarkable features arising from the conjunction of the following elements: 

\noindent {\it 1 - Duality of typical and rare events}. 
The distinction between typical and rare events is relative to a
given statistical ensemble (the prior), reflecting the conditions of observation. 
Of particular current interest, in many different contexts ranging from classical~\cite{derrida,Nemoto,classical2} to
quantum~\cite{quantum}, is the task of constructing
a dual ensemble in which an event, rare in the prior, becomes typical
in the dual one. This kind of problem is addressed and solved in the framework of large 
deviation theory~\cite{Touchette}.

\noindent {\it 2 - Condensation of fluctuations}. 
Condensation transitions as average or typical properties are ubiquitous and
familiar phenomena, much studied, both in equilibrium and out of equilibrium, since a
long time. Suffice it to quote the two celebrated examples of the condensation of supersaturated vapor
in real space and of the Bose-Einstein condensation (BEC) in momentum space~\cite{Huang}.
Much less familiar, and of more recent interest, is the concept of condensation of fluctuations,
that is of condensation as a rare event~\cite{schutz,prec,Nossan,PREZ}. 
These two instances of condensation, on average and of fluctuations,
are the two facets of a unique phenomenon, related via the above mentioned ensemble duality and
corresponding to different observation protocols.

\noindent {\it 3 - Condensation triggered by equipartition breaking}.
We shall restrict to condensation transitions occurring through a mechanism
{\it \`a la} BEC. Whether such a mechanism can be implemented or not depends on the observable of
interest and on the chosen statistical ensemble. We shall be chiefly interested in the energy
as a macrovariable and we shall see that, for the mechanism {\it \`a la} BEC to operate,
the equipartition of energy must be violated. Hence, condensation of energy fluctuations arises 
{\it exclusively} as an out of equilibrium phenomenon {\it and} in the transient regime.
Condensation of fluctuations arising both in and out of equilibrium have been recently
analyzed in Ref.~\cite{PREZ}.

From the combined occurrence of these three elements there arises a rich and interesting
phenomenology, which will be studied as the system relaxes from one equilibrium state to another.
Specifically, we shall consider an instantaneous temperature quench: The system
is initially prepared in equilibrium at the temperature $T_I$ and, then, at the time
$t=0$ is put in contact with a new thermal reservoir at the lower temperature $T_F$, 
producing the relaxation toward the new equilibrium state.
We shall consider an extended system with a
phase diagram containing a critical point at the temperature $T_C$.
Just to fix the ideas, we may think of $T_C$ as the Curie temperature of a ferromagnet, which
is paramagnetic above $T_C$ and ferromagnetic below $T_C$, but the considerations we shall make
hold in general. The reason for studying such a process is that if the quench is made from
$T_I$ to $T_F$, both above $T_C$, then there is a finite equilibration time 
allowing to overview the full evolution of fluctuations from initial equilibrium, to intermediate
off-equilibrium behavior and again to equilibrium in the final state. Instead,
if $T_I$ is taken above and $T_F$ below $T_C$, the system undergoes phase-ordering~\cite{Bray},
where the equilibration time scales like the size of the
system. Thus, a system of thermodynamic size remains, for all practical purposes, permanently
out of equilibrium and, as we shall see, there arise important qualitative differences
in the behavior of fluctuations~\cite{crisanti,ciliberto,prec}.


As observables, in addition to the total energy chosen as a one-time macrovariable, we shall also consider
the heat exchanged with the cold reservoir in a time interval $(t_w,t)$, with $t_w < t$, as a
two-times macrovariable. The model system will be treated analytically, via a mean-field theory
which allows to effectively decouple the degrees of freedom. 
This is an important feature, since, after introducing the
notion of effective temperature per degree of freedom, 
the transition from equilibrium to off-equilibrium can be
characterized as an equipartition breaking process. 
This goes as follows. In the initial equilibrium state all degrees
of freedom are at the same temperature $T_I$ and equipartition holds. As soon as relaxation begins,
immediately after the quench, the spectrum of effective temperatures is no more flat
and equipartition is broken~\cite{rondoni}. The amount of equipartition breaking,
encoded in the deviation from flatness of the temperature spectrum, quantifies
the amount of departure from equilibrium. Eventually, if the system equilibrates
equipartition is restored, while if the system does not equilibrate, as in
the quench to below $T_C$, equipartition remains permanently broken according to
a pattern characteristic of the asymptotic regime.

A remarkable consequence of equipartition breaking, as mentioned above, is the appearance
of the condensation of fluctuations, when
the temperature spectrum develops a minimum or a maximum at zero wave vector.
Then, the entire amount of the fluctuation exceeding a certain critical threshold
is contributed by the zero-wave-vector degree of freedom, via a 
mechanism whose mathematics is the same of BEC.
The nature of this condensation phenomenon will be analyzed in detail.
Here we stress, first of all, that the analogy with BEC is only formal, since 
BEC is about an average equilibrium property, while condensation of fluctuations
is about rare events off-equilibrium. Secondly, condensation phenomena have been
recently studied in a variety of different contexts, ranging from the realm of statistical physics (in and
out of equilibrium, classical~\cite{BK,castellano,ZPRreview} and quantum~\cite{Gambassi}) 
to problems of interest in economics
and information theory~\cite{Marsili}. However, most of these works deal with identically distributed
variables and with stationary states. The specificity of what we do in the present work
is that variables are not identically distributed and in the non
stationary nature of the relaxation process. Both of these features are essential 
for the appearance of the breaking of equipartition, which is at the basis of the condensation phenomenon.
 
The paper is organized as follows: In Sec.~\ref{EH} we introduce the general concepts needed
in the study of the fluctuations of macrovariables, with particular attention to the ensemble theory structure
underlying the dichotomy typical vs rare events. The ideal Bose gas is briefly worked out
as an example and as a reminder of BEC. In Sec.~\ref{largeN}, after presenting the
large $N$ model and its exact solution, we set up the formal apparatus for the study of
fluctuations of energy and heat. The actual study of fluctuations is carried out in Sec.~\ref{para}
and Sec.~\ref{ferro}, devoted, respectively, to the relaxation within the paramagnetic phase
and from above to below $T_C$. Concluding remarks are made in Sec.~\ref{conclusions}.

\section{Fluctuations of a random variable}
\label{EH}

In this section we introduce the general concepts of fluctuations
of a macrovariable, of typical and rare events, and of
ensemble duality. 

Consider a generic prior probability distribution $P(\varphi,J)$ of elementary events $\varphi$ 
in a phase space $\Omega$ and with control parameters $J$. 
The probability of a fluctuation $M$ of a random variable
${\cal M}(\varphi)$ is given by
\be 
P(M,J) = \int_{\Omega} d\varphi \, P(\varphi,J)\delta (M-{\cal M}(\varphi)).
\label{gen.3}
\ee
Most of the times this is just a formal expression, since the restriction of the
integration domain, due to the $\delta$ function, makes the calculation
impracticable. The phase space can be maintained intact by 
introducing the integral representation of the $\delta$ function
\be
\delta(x) = \int_{\alpha -i\infty}^{\alpha +i\infty}\frac{dz}{2\pi i} \, e^{-zx},
\label{gen.4}
\ee
which, then, shifts the problem to the computation of the Fourier transform
\be
P(M,J) =  \int_{\alpha - i\infty}^{\alpha + i\infty} \frac{dz}{2\pi i} \, e^{-zM} K_{\cal M}(z,J),
\label{gen.5}
\ee
where
\be
K_{\cal M}(z,J) = \langle e^{z{\cal M}(\varphi)} \rangle
\label{gen.6}
\ee
is the moment generating function of ${\cal M}$. 
The brackets $\langle \cdot \rangle$ denote the average in the prior ensemble.
  
If the system is extended and ${\cal M}(\varphi)$ is an extensive macrovariable,
for large volume Eq.~(\ref{gen.5}) can be rewritten as
\be
P(M,J,V) =  \int_{\alpha - i\infty}^{\alpha + i\infty} \frac{dz}{2\pi i} \, e^{-V[zm + \lambda_{\cal M}(z,J)]},
\label{gen.6}
\ee
where the $V$ dependence is made explicit, $m$ is the density $M/V$ and
\be
-\lambda_{\cal M}(z,J) = \frac{1}{V}\ln K_{\cal M}(z,J,V)
\label{gen.7bis}
\ee 
is the volume independent scaled cumulant generating function. 
Carrying out the integration by the saddle point method,
the large deviation principle is obtained
\be
P(M,J,V) \sim e^{-VI_{\cal M}(m,J)},
\label{gen.8}
\ee
with rate function 
\be
I_{\cal M}(m,J) = z^*m + \lambda_{\cal M}(z^*,J), 
\label{gen.9}
\ee
where $z^*(m,J)$ is the solution, supposedly unique, of the saddle point equation
\be
\frac{\partial}{\partial z} \lambda_{\cal M}(z,J) = -m.
\label{gen.9bis}
\ee
From the above algebra follows the basic result of large deviation theory~\cite{Touchette} that
$I_{\cal M}(m,J)$ and $\lambda_{\cal M}(z,J)$ form a pair of Legendre transforms.

\vspace{1cm}

\noindent {\it Typical and rare events}

\vspace{1cm}

The large deviation principle implies that the probability of $m$ is 
concentrated about the most probable value $m^*$,
identified by the condition $\partial_m I_{\cal M}(m,J) = z^*(m,J) = 0$.
This allows to discriminate between typical and rare events.
Typical are events in the immediate neighborhood $m^*$.
All other outcomes of $m$, lying on the tails of the rate function, 
are rare events, since they do occur with an exponentially low probability.
Clearly, the qualification of an outcome as typical or rare, is relative to
the given prior distribution, that is, to the given preparation protocol, specified
by the set of control parameters $J$. For instance, for a thermodynamic system
$J$ contains the list of the conserved extensive quantities and of the applied
intensive fields, like temperature, pressure and so on.
Then, one of the key features of the rate function is that it may serve
a twofold purpose: either as the just specified quantifier of the rarity of a large deviation in the prior,
or as the prescription of how to change the observation protocol, 
in order to render typical an event which is rare in the prior.
In order to understand the latter statement, let us rewrite Eq.~(\ref{gen.9bis}) as
\be
m= \frac{1}{V} \langle {\cal M} \rangle_{z^*},
\label{GELLIS.01}
\ee
where $\langle \cdot \rangle_z$ stands for the average with respect to the distribution
\be
\mathbb{P}(\varphi,z,J,V) = \frac{1}{K_{\cal M}(z,J,V)} P(\varphi,J,V) \, e^{z{\cal M}(\varphi)}.
\label{Du.2}
\ee
Then, the meaning of Eq.~(\ref{GELLIS.01}) is that
the event $m$ becomes typical in the statistical ensemble obtained
by imposing the exponential bias $e^{z^*{\cal M}(\varphi)}$ on the prior, which means that the preparation
protocol must be changed by putting the system in contact with a reservoir of the observable ${\cal M}$ and
by fixing the intensive parameter $z$, conjugated to ${\cal M}$, to the value $z^*(m,J)$ such that
condition~(\ref{GELLIS.01}) is fulfilled.
Consequently, $\lambda_{\cal M}(z,J)$ plays the role of the
``thermodynamic potential'' in the biased statistical ensemble and, the rate function $I_{\cal M}(m,J)$,
as the Legendre transform of $\lambda_{\cal M}(z,J)$, plays the role of the thermodynamic
potential of the system prepared with yet another protocol, that is 
by constraining rigidly $\frac{1}{V}{\cal M}$ to take the value
$m$, which is nothing but the obvious way to render typical the event $m$. The statistical ensemble
of the latter protocol, to be referred to as the constrained ensemble,
is obtained by projecting the prior on the subset of events satisfying the imposed constraint 
\be
\mathfrak{P}(\varphi,M,J,V) = \frac{1}{P(M,J,V)} P(\varphi,J,V)\delta (M-{\cal M}(\varphi)).
\label{gen.2}
\ee

Summarizing, the twofold role of the rate function brings into the picture
{\it three} statistical ensembles: the prior on one side, denoted by $P$, and the biased and the constrained ensembles
on the other, denoted by $\mathbb{P}$ and $\mathfrak{P}$, respectively. 
The remarkable aspect of these formal relations is in the predictive power
of $I_{\cal M}(m,J)$, in the sense that by observing fluctuations in the prior we can predict
what will happen in the biased or in the constrained ensemble and vice versa, much in the same
way as it works with the fluctuation dissipation relations~\cite{Nemoto}. When $I_{\cal M}(m,J)$ 
is obtained from the biased or constrained ensemble, it plays the role of
a {\it fluctuation theory}. Conversely, the observation of fluctuations is crucial
when constraints or biasing fields cannot be implemented in practice, as we shall
see in the following. Lastly, the duality of $I_{\cal M}(m,J)$
becomes of particular interest when singularities appear~\cite{Bunin}. Then,
looking at $I_{\cal M}(m,J)$ as a thermodynamic potential, singular behavior
is readily interpreted as symptomatic of a phase transition~\cite{Bertini,Nossan}, whose counterpart in the context
of the prior ensemble is the novel and unfamiliar phenomenon of 
a phase transition in the behavior of fluctuations.
This is all the more interesting because condensation of the fluctuations may occur,
as we shall see, even in non interacting systems, which cannot sustain condensation as an average
thermodynamic property. 

As anticipated in the Introduction,
the study of fluctuations condensation is a central theme in this paper. 
However, before addressing the problem off-equilibrium, 
in the following subsection the concept of duality will be illustrated in the usual framework
of equilibrium statistical mechanics.

\subsection{Ideal Bose gas in thermal equilibrium}
\label{ideal}

Let us consider an ideal gas of bosons in a box of volume $V=L^{d}$,
where $d$ is the space dimensionality of the system.
The microstates are the sets of occupation numbers $\varphi= \{ n_{\vec p} \}$ of the 
single particle momentum eigenstates
$\vec p = \hbar \vec k$, where, imposing periodic boundary conditions, the allowed
wave vectors values are
\be
\vec k = \frac {2\pi}{L} \vec m, \;\;\; m_i=0,\pm 1, \pm 2,...
\label{IBG.1bis}
\ee
In the following we shall take $\hbar =1$ and, therefore, $\vec p = \vec k$.
The extensive macrovariables of interest are the energy, which is separable
\be
{\cal H}(\varphi) = \sum_{\vec k} {\cal H}_{\vec k}(n_{\vec k}),
\label{IBG.2}
\ee
with ${\cal H}_{\vec k}(n_{\vec k}) =  n_{\vec k}\epsilon_k$ and the number function
\be
{\cal N}(\varphi) = \sum_{\vec k} n_{\vec k}.
\label{IBG.3}
\ee
We assume that the single particle dispersion relation is of power law form $\epsilon_k = ak^{\alpha}$,
where $a$ is a proportionality constant. For instance, for photons $a=c$, velocity of light, and $\alpha = 1$, while
for particles with mass $m$, $a=1/(2m)$ and $\alpha = 2$.

When the system is in equilibrium with a thermal bath at the temperature $\beta^{-1}$
(taking $k_B=1$, as in the rest of the paper) and with a particle reservoir at the chemical potential
$\mu$, the prior is the grand canonical ensemble
\be
P(\varphi,\beta,\mu,V) = \frac{1}{Z_{\rm gc}(\beta,\mu,V)} e^{-\beta [ {\cal H}(\varphi)-\mu{\cal N}(\varphi)]},
\label{IBG.50}
\ee
where
\be
Z_{\rm gc}(\beta,\mu,V) = \int_{\Omega(V)} d\varphi \, e^{-\beta [ {\cal H}(\varphi)-\mu{\cal N}(\varphi)]}
\label{IBG.51}
\ee
is the grand partition function and $\Omega(V)$ is the phase space, 
with no restriction on the total particle number. 

We shall now look at the the fluctuations of
energy and number in two particular cases.

\subsubsection{Energy fluctuations with $\beta =0$}

In the limit of infinite temperature ($\beta =0$), the prior ensemble becomes
uniform
\be
P(\varphi,V) = \frac{1}{|\Omega(V)|},
\label{IBG.52}
\ee
where $|\Omega(V)|$ is the phase space volume. This case is treated
also in the review by Touchette~\cite{Touchette}.
According to Eq.~(\ref{gen.3}), the total energy fluctuates with probability
\begin{eqnarray}
P(E,V) & = & \int_{\Omega(V)} d\varphi \, P(\varphi,V)\delta (E-{\cal H}(\varphi)) \nonumber \\
& = & \frac{|\Omega(E,V)|}{|\Omega(V)|},
\label{IBG.54}
\end{eqnarray}
where $\Omega(E,V) = \{\varphi|{\cal H}(\varphi) = E \}$ is the subset of $\Omega(V)$
satisfying the energy constraint. Denoting by $e = E/V$ the energy density,
$P(E,V)$ obeys the large deviation principle with the rate function
\be
I_{\cal H}(e) = s_{\rm act} - s(e),
\label{IBG.56}
\ee
where $s_{\rm act} = \frac{1}{V} \ln |\Omega(V)|$ is the entropy density in the actual state of the
system and $s(e) = \frac{1}{V} \ln |\Omega(E,V)|$ is the entropy density that the system would have
if the system were isolated with total energy $E$. Therefore, the constrained ensemble $\mathfrak{P}(\varphi,E,V)$
is the microcanonical ensemble corresponding to the thermodynamic state $(E,V)$ and
Eq.~(\ref{IBG.56}) is nothing but Einstein fluctuation theory.

The formal structure is completed by the cumulant generating function
\be
\lambda_{\cal H}(z) = -\frac{1}{V}\ln \langle e^{z{\cal H}} \rangle = -\frac{1}{V}\ln\frac{Z_{\rm gc}(-z,0,V)}{Z_{\rm gc}(0,0,V)} 
\label{IBG.57}
\ee  
and by the biased ensemble
\be
\mathbb{P}(\varphi,z,0,V) = \frac{1}{Z_{\rm gc}(-z,0,V)}  e^{z{\cal H}(\varphi)},
\label{IBG.60}
\ee
which is the grand canonical ensemble with $\beta = -z$ and $\mu=0$. Recalling 
that the grand partition function satisfies
\be
-\frac{1}{V}\ln Z_{\rm gc}(\beta,\mu,V) = \beta \overline{e}(\beta,\mu)
-s(\overline{e}) -\beta \mu \overline{\rho}(\beta,\mu),
\label{GCan.1}
\ee
where $\overline{e}(\beta,\mu)$ and $\overline{\rho}(\beta,\mu)$ are the average energy and number
densities, and that the Massieu potential $\mathfrak{s}(\beta) = s(e) - \beta e$  is the Legendre transform
of entropy with respect to energy~\cite{Callen},
from Eq.~(\ref{IBG.57}) follows
\be
\lambda_{\cal H}(z) = s_{\rm act} - \mathfrak{s}(-z),
\label{GCan.3}
\ee
as it should be, since $\lambda_{\cal H}(z)$ is the Legendre transform of $I_{\cal H}(e)$.

\subsubsection{Number fluctuations with $\mu = 0$}
\label{BEC}

Before looking at fluctuations, let us recall that the average density 
$\overline{\rho}(\beta,\mu)= \langle {\cal N} \rangle /V$ in
the grand canonical ensemble is given by the sum of the average occupation numbers
\be
\overline{\rho} = \frac{1}{V} \sum_{\vec k} \langle n_{\vec k} \rangle =
\frac{1}{V} \sum_{\vec k} \frac{1}{e^{\beta (\epsilon_k-\mu)}-1}
\label{dens.1}
\ee
or, taking $V$ large and transforming the sum into an integral, by
\be
\overline{\rho} = \frac{\Upsilon_d}{(2\pi)^d}\int_0^{\infty} d k \, \frac{k^{d-1}}{e^{\beta (\epsilon_k-\mu)} -1},  
\label{dens.2}
\ee
where $\Upsilon_d = 2\pi^{d/2}/\Gamma (d/2)$ is the $d$-dimensional solid angle
and $\Gamma$ is the Euler gamma function. Keeping $\beta$ fixed, 
$\overline{\rho}$ grows monotonically as $\mu$ increases from $-\infty$, reaching the upper bound at $\mu = 0$
\be
\overline{\rho}_C(\beta) = \frac{\Upsilon_d}{(2\pi)^d}\int_0^{\infty} d k \, \frac{k^{d-1}}{e^{\beta \epsilon_k} -1},
\label{dens.3}
\ee
which is infinite for $d \leq \alpha$ and is finite for $d > \alpha$. In the latter case, this finite
upper bound has two different meanings,
depending on whether $\overline{\rho}$ is conserved or not. For photons
or phonons, which live on the $\mu=0$ axis without number conservation, $\overline{\rho}_C(\beta)$ simply gives 
the temperature dependence of the average density. 
Instead, if the density is conserved as for atoms, $\overline{\rho}_C(\beta)$
is the critical value beyond which BEC occurs. In the standard textbook treatment~\cite{Huang},
this is obtained by separating the first term from the sum
in Eq.~(\ref{dens.1}) and by rewriting it as
\be
\overline{\rho} = \frac{1}{V} \frac{1}{(e^{-\beta \mu}-1)} + \overline{\rho}_C.
\label{dens.4}
\ee
Hence, for $\overline{\rho} > \overline{\rho}_C$ one has 
$\langle n_0 \rangle = V(\overline{\rho} - \overline{\rho}_C)$, which
implies $-\mu \sim 1/V$.

Let us now come to the number fluctuations when $\beta$ is finite and $\mu=0$.
The probability that ${\cal N}$ takes the value $N$ is given by
\begin{eqnarray} 
P(N,\beta,0,V) & = & \int_{\Omega(V)} d\varphi \, P(\varphi,\beta,0,V)\delta (N-{\cal N}(\varphi)) \nonumber \\
& = & \frac{Z_{\rm  c}(\beta,N,V)}{Z_{\rm gc}(\beta,0,V)},
\label{mu.1}
\end{eqnarray}
where
\be
P(\varphi,\beta,0,V) = \frac{1}{Z_{\rm gc}(\beta,0,V)} e^{-\beta {\cal H}(\varphi)}
\label{mu.0}
\ee
is the prior, that is the grand canonical ensemble with $\mu = 0$, while
$Z_{\rm  c}(\beta,N,V)$ in the right hand side of Eq.~(\ref{mu.1}) is the {\it canonical} partition function 
with the number of particles fixed
to $N$. Exponentiating and denoting by $\rho=N/V$ the {\it fluctuating} density, 
the large deviation principle follows with rate function 
\be
I_{\cal N}(\beta,\rho) = \mathfrak{s}_{\rm act}(\beta) - \mathfrak{s}(\beta,\rho),
\label{mu.3}
\ee
where $\mathfrak{s}_{\rm act}(\beta)$ and $\mathfrak{s}(\beta,\rho)$ are Massieu potentials
in the actual state and in the state $(\beta,\rho)$. Therefore, 
the constrained ensemble $\mathfrak{P}(\varphi,\beta,\rho)$  is the canonical ensemble in the state $(\beta,\rho)$.

It is not difficult to check that the same result is obtained by following the route of
Eq.~(\ref{gen.9}), which yields
\be
I_{\cal N}(\beta,\rho) = z^*(\beta,\rho) \rho 
-\frac{1}{V}\ln \left [ \frac{Z_{\rm gc}(\beta,\mu(z^*),V)}{Z_{\rm gc}(\beta,0,V)} \right ],
\label{IBG.5}
\ee
where $\mu(z^*) = \beta^{-1} z^*$ is the chemical potential determined
by the saddle point equation
\be
\rho = \frac{1}{V} \langle {\cal N} \rangle_{z}.
\label{IBG.32}
\ee

\begin{figure}
\begin{center}
\epsfysize=6.0cm \epsffile{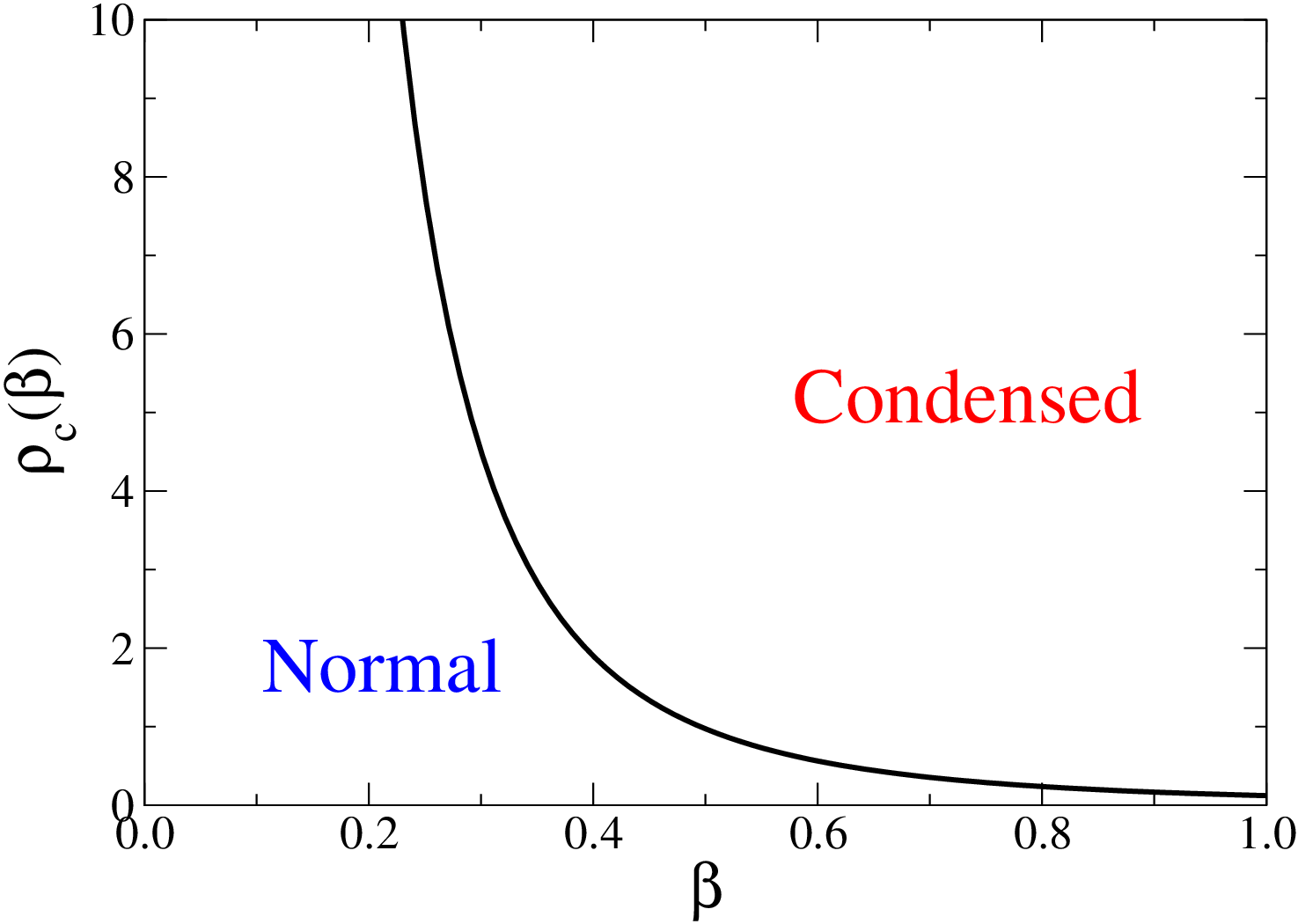} 
\caption{Boson fluctuations phase diagram for $d=3$, $a=1$ and $\alpha=1$.
See Eq.~(\ref{IBG.41}).}
\label{BEC}
\end{center}
\end{figure}

\noindent This is the key equation in the present discussion, since it is of the same form as 
Eq.~(\ref{dens.1}), except that in the left hand side now there appears the chosen value
of the fluctuating density. First of all, this means that the large deviation $\rho$,
in the prior, becomes a typical event in the biased ensemble $\mathbb{P}(\varphi,\beta,z^*,V)$, 
which is the grand canonical
ensemble with chemical potential fixed by $z^*$. According to the previous
discussion of BEC, $z^*$ is driven to zero for $\rho \geq \overline{\rho}_C$ and, as it is evident
from Eq.~(\ref{IBG.5}), there follows that also $I_{\cal N}(\rho,\beta)$ vanishes
for $\rho \geq \overline{\rho}_C$. Consequently, $P(N,\beta,0,V)$ becomes uniform
for $\rho \geq \overline{\rho}_C$, which is the manifestation of the condensation transition
in the prior or the condensation of the fluctuations~\cite{Gambassi}.
In this context, the critical value~(\ref{dens.3}) acquires the different meaning
of the threshold beyond which fluctuations condense. Plotting
$\overline{\rho}_C(\beta)$ in the $(\rho,\beta)$ plane, the phase diagram is obtained (see Fig.~\ref{BEC})
with the phases standing for different behaviors of the fluctuations.
With a linear dispersion relation, like for photons or phonons,
from Eq.~(\ref{dens.3})
with $d=3$ and $a=1$ we have
\be
\overline{\rho}_C(\beta) = \frac{\zeta(3)}{\pi^2 \beta^3},
\label{IBG.41}
\ee
where $\zeta(3) = 1.202..$ is the Riemann zeta function (see Fig.~\ref{BEC}).

\section{The large $N$ model}
\label{largeN}

After the equilibrium examples, let us come to the central theme of the paper, which
is the study of the fluctuations in a temperature quench. The system undergoing the quench
is the $N$ vector model. This is a classical system, 
defined by the Ginzburg-Landau energy functional~\cite{Goldenfeld}
\be
{\cal H}_N[\vec{\varphi}] =
\frac{1}{2} \int_V d \vec x \, \left [(\nabla \vec {\varphi})^2 + r \vec {\varphi}^2
+ \frac{g}{2N} (\vec {\varphi}^2)^2 \right ],
\label{Gin.1}
\ee
where $\vec {\varphi}(\vec x) = [\varphi_{\alpha}(\vec x)]$, with $\alpha=1,..N$, is a configuration
of the $N$ component vectorial order parameter and $g > 0$. As it is well known, this is the basic generic model for the
study of systems with continuos symmetry, but for simplicity we shall adopt the magnetic language. 
If $r$ is non negative the local potential
is of the one-well type and the model describes only
the paramagnetic phase, irrespective of the value of the temperature. 
Instead, if $r$ is negative, the local potential is of the mexican hat type and
there exists a critical point separating an high temperature paramagnetic phase from a
low temperature ferromagnetic phase.

The dynamics, without conservation of the order parameter, are governed by the overdamped Langevin
equation~\cite{HH}
\be
\dot{\varphi}_{\alpha} = -\frac{\delta}{\delta \varphi_{\alpha}}{\cal H}_N[\vec \varphi] + \eta_{\alpha},
\label{LG.1}
\ee
where $\vec{\eta}(\vec x, t)$ is the white Gaussian noise generated by the cold reservoir, 
with zero average and correlators
\be
\langle \eta_{\alpha}(\vec x,t)\eta_{\beta}(\vec {x}^{\prime},t^{\prime}) \rangle = 
2T_F \delta_{\alpha \beta}\delta(\vec x -\vec {x}^{\prime})   \delta(t-t^{\prime}).
\label{LG.2}
\ee

One of the ways to treat the model perturbatively is to take $1/N$ as the expansion parameter
when $N$ is large.
To lowest order (large $N$ limit) the model becomes exactly solvable~\cite{Goldenfeld,CA}.
In order to see how this comes about, consider the explicit form of the equation of motion
\be
\dot{\varphi}_{\alpha} = -\left ( -\nabla^2 + r 
+ \frac{g}{N} \vec {\varphi}^2 \right ) \varphi_{\alpha}   + \eta_{\alpha}.
\label{Gin.2}
\ee
Then, as $N$ becomes large, the nonlinear term is self-averaging
\be
\lim_{N \to \infty}\frac{1}{N} \vec {\varphi}^2 = \langle \varphi_{\alpha}^2(\vec x,t) \rangle = S(t)
\label{Gin.3}
\ee
and, for an homogeneous system, $S(t)$ is independent of the component label and 
of the space position. With this replacement, the equation of motion becomes formally linear
\be
\dot{\varphi}_{\alpha} = -\left [ -\nabla^2 + r + gS(t) \right ] \varphi_{\alpha}   + \eta_{\alpha}
\label{Gin.4}
\ee
and can be regarded as arising from Eq.~(\ref{LG.1}) with
the time dependent effective energy functional
\be
{\cal H}_{\infty}[\vec \varphi,t] = \sum_{\beta=1}^{\infty}{\cal H}_{\rm eff}[ \varphi_{\beta},t],
\label{Gin.4bis}
\ee
where
\be
{\cal H}_{\rm eff}[ \varphi_{\beta},t] =
\frac{1}{2} \int_V d \vec x \, \left \{ (\nabla  \varphi_{\beta})^2 + [r + g S(t)] \varphi_{\beta}^2 \right \}.
\label{Gin.5}
\ee
That is, by taking the large $N$ limit the original system ${\cal H}_N[\vec \varphi]$ has been replaced by the sum of
infinitely many independent replicas of a new system, described by ${\cal H}_{\rm eff}[ \varphi_{\beta},t]$,
in which the original coupling of components has generated the mean-field mass renormalization through $S(t)$.
Then, from now on we shall work with the single replica, dropping the component label.
Notice that, due to the binding local mexican hat potential, ${\cal H}_N$ is bounded
below, while ${\cal H}_{\rm eff}$ is unbounded below, because the curvature of the time dependent local harmonic
potential can become negative, as we shall see in Sec.~\ref{ferro}.  

Dealing with a formally linear problem, the dynamics can be diagonalized by Fourier transformation.
However, some care is needed
in the identification of the independent variables, keeping in mind that the Fourier components
$\varphi_{\vec k} = \int_V d\vec x \, \varphi(\vec x) e^{i\vec k \cdot \vec x}$ are complex.
Adopting periodic boundary conditions, the allowed wave vectors values are those specified
in Eq.~(\ref{IBG.1bis}).
Let us denote by ${\cal B}$ the set of all these wave vectors with magnitude smaller than
an ultraviolet cutoff $\Lambda$, due to the existence of a microscopic length scale
in the problem, like an underlying lattice spacing. 
Since the reality of $\varphi(\vec x)$ requires $\varphi_{-\vec k}=\varphi_{\vec k}^*$,
the independent variables are $\varphi_0$ 
and the set of pairs $ \{\mathbb{R}e \, \varphi_{\vec k}, \; \mathbb{I}m \, \varphi_{\vec k} \}$
with $\vec k \in {\cal B}_+$, where ${\cal B}_+$ is a half of ${\cal B}$.
More precisely, if ${\cal B}_-$ is the set obtained by reversing all the wave vectors in ${\cal B}_+$,
then ${\cal B}_+$ is such that ${\cal B}_+  \cap {\cal B}_- = \emptyset$ and 
${\cal B}_+  \cup {\cal B}_- = {\cal B}-\{ \vec 0 \}$.
However, rather than working with ${\cal B}_+$, it is more convenient to
let $\vec k$ to vary over the whole of ${\cal B}$ by taking as independent real variables
\be
x_{\vec k}  = \left \{ \begin{array}{ll}
         \varphi_0 ,\;\; $for$ \;\; \vec k = 0  ,\\
         \mathbb{R}e \, \varphi_{\vec k},\;\; $for$ \;\; \vec k \in {\cal B}_+, \\
         \mathbb{I}m \, \varphi_{\vec k},\;\; $for$ \;\; \vec k \in {\cal B}_-.
        \end{array}
        \right .
        \label{frrr.3}
        \ee
With this convention, from Eq.~(\ref{Gin.4}) we get
the equations of motion for a set of independent Brownian oscillators
\be
\dot{x}_{\vec k}(t) = -\omega_k(t) x_{\vec k}(t) + \zeta_{\vec k}(t),
\label{FOUR.2}
\ee 
\be
\omega_k(t) = [k^2 + r + gS(t)],
\label{Gin.6}
\ee
where $\zeta_{\vec k}(t)$ is related to the Fourier transform $\eta_{\vec k}(t)$ of the thermal noise
in Eq.~(\ref{Gin.4}) by the analogue of Eq.~(\ref{frrr.3})
\be
\zeta_{\vec k}(t)  = \left \{ \begin{array}{ll}
         \eta_0(t) ,\;\; $for$ \;\; \vec k = 0  ,\\
         \mathbb{R}e \, \eta_{\vec k}(t) ,\;\; $for$ \;\; \vec k \in {\cal B}_+, \\
         \mathbb{I}m \, \eta_{\vec k}(t) ,\;\; $for$ \;\; \vec k \in {\cal B}_-.
        \end{array}
        \right .
        \label{frrr.3bis}
        \ee
It is straightforward to check that this is also a zero average Gaussian white noise, with correlators 
\be
\langle \zeta_{\vec k}(t) \zeta_{\vec{k}^{\prime}}(t^{\prime}) \rangle  = 2T_{F,k} V \delta_{\vec k,\vec{k}^{\prime}}\delta(t-t^{\prime}),
\label{FOUR.8}
\ee
where
\be
T_{F,k} = \frac{T_F}{2\theta_k}
\label{finT.1}
\ee
and $\theta_k$ is the Heaviside step function with $\theta_0=1/2$.
The energy functional~(\ref{Gin.5}) takes the separable form
\be
{\cal H}_{\rm eff}[\mathbf{x},t] = \sum_{\vec k}{\cal H}_{\vec k}(x_{\vec k},t),
 \label{GMD.2bis}
\ee
with
\be
{\cal H}_{\vec k}(x_{\vec k},t) = \frac{1}{V}\theta_k \omega_k(t) x^2_{\vec k},
\label{FOUR.9}
\ee
where $\mathbf{x}$ stands for the whole set $\{x_{\vec k}\}$.

Integrating Eq.~(\ref{FOUR.2}) from $t_w$ to $t > t_w$, with the initial condition $x_{w,\vec k}$,
we obtain
\be
x_{\vec k}(t) = G_k(t,t_w)x_{w,\vec k} + \int_{t_w}^t dt^{\prime} \,G_k(t,t^{\prime}) \zeta_{\vec k}(t^{\prime}),
\label{newver.1}
\ee
where 
\be
G_k(t,t^{\prime})  = \exp \left \{-\int_{t^{\prime}}^t ds \, \omega_k(s)  \right \}
\label{FOUR.7}
\ee
is the Green's function of Eq.~(\ref{FOUR.2}). Taking averages with respect to the noise, the average
and the variance of $x_{\vec k}(t)$ are given by
\be
\overline{x}_{\vec k}(t,t_w) = G_k(t,t_w)x_{w,\vec k} 
\label{FOUR.5}
\ee
and
\be
\sigma_{k}(t,t_w) = \overline{[x_{\vec k}(t) -\overline{x}_{\vec k}(t,t_w)]^2} =      
2T_{F,k} V \int_{t_w}^{t} dt^{\prime} \, G^2_k(t,t^{\prime}). 
\label{FOUR.6}
\ee
Therefore, due to the linearity of the equation of motion and to the Gaussian statistics
of the noise, the transition probability is given by
\be
R_{\vec k}(x_{\vec k},t|x_{w,\vec k},t_w) = \frac{1}{\sqrt{2 \pi \sigma_{k}(t,t_w)}}
\exp \left \{-\frac{[x_{\vec k} - \overline{x}_{\vec k}(t,t_w)]^2}{2  \sigma_{k}(t,t_w)} \right \}.
\label{FOUR.4}
\ee
From this, it is simple to derive the autocorrelation function
\be
\langle x_{\vec k}(t) x_{\vec k}(t_w) \rangle = G_k(t,t_w) 
\left [G^2_k(t_w,0)\langle x^2_{\vec k}(0) \rangle  + \sigma_{k}(t_w,0) \right ],
\label{FOUR.70}
\ee
where we have assumed a symmetrical initial condition, implying
$\langle x_{\vec k}(t) \rangle =0$ for all times. In the above formulas the overline denotes an average
with respect to the thermal noise, while the angular brackets are for averages with respect to 
all sources of noise: thermal bath and initial conditions.

\subsection{Exact solution}

The actual solution of the model~\cite{CA,Godreche} requires the determination of $S(t)$,
which can be rewritten as 
\be
S(t) = \frac{1}{V} \sum_{\vec k} C_{\vec k}(t),
\label{Gin.8}
\ee
where
\be
C_{\vec k}(t) = 2\theta_k \langle x^2_{\vec k}(t) \rangle 
\label{Gin.9}
\ee
is the structure factor, namely the Fourier transform of the equal times real space
correlation function $C(\vec x - \vec x^{\prime},t) =  \langle \varphi(\vec x,t)\varphi(\vec x^{\prime},t) \rangle$.
Integrating Eq.~(\ref{FOUR.2})
and using Eq.~(\ref{FOUR.8}), one finds
\be
C_{\vec k}(t) = G^2_k(t,0)C_{\vec k}(0) + 2T_F \int_0^t dt^{\prime} \, 
G^2_k(t,t^{\prime}).
\label{Gin.16}
\ee
Therefore, using 
\be
-\frac{\partial}{\partial t} \ln G_0(t,0) =  \omega_0(t),
\label{frrr.5}
\ee
\be
G_k(t,t^{\prime}) = e^{-k^2(t-t^{\prime})} G_0(t,t^{\prime})
\label{frrr.6}
\ee
and inserting Eq.~(\ref{Gin.16}) into Eq.~(\ref{Gin.8}),
the problem is closed by the integro-differential equation for $G_0(t,t_w)$
\be
-\frac{\partial}{\partial t} \ln G_0(t,0) = r +g\frac{1}{V} \sum_{\vec k}
\left [ G^2_k(t,0)C_{\vec k}(0) + 2T_F \int_0^t dt^{\prime} \, 
G^2_k(t,t^{\prime}) \right ].
\label{Gin.12}
\ee
For future reference, we mention that from Eq.~(\ref{Gin.16})
it is straightforward to verify that the structure factor satisfies the equation of motion
\be
\frac{\partial}{\partial t} C_{\vec k}(t) = -2 \omega_k(t) C_{\vec k}(t) + 2T_F.
\label{Gin.10}
\ee

\subsection{Statics}

The equilibrium properties of the large $N$ model are well known~\cite{castellano,Goldenfeld}. 
Here, we list a few results needed in the following.
From the separable form~(\ref{GMD.2bis}) of the energy functional follows that the Gibbs state is factorized
\be
P_{\rm eq}[\mathbf{x},\beta] = \prod_{\vec k} P_{{\rm eq},\vec k}(x_{\vec k},\beta),
\label{Equil.1}
\ee
with
\be
P_{{\rm eq},\vec k}(x_{\vec k},\beta)  =  Z^{-1}_{{\rm eq},\vec k}(\beta) e^{-\beta{\cal H}_{{\rm eq},\vec k}(x_{\vec k})},
\label{Equil.2}
\ee
\be
Z_{{\rm eq},\vec k}(\beta)  = \sqrt{\frac {\pi V}{\beta \theta_k \omega_{{\rm eq},k}}}.
\label{Equil.3}
\ee
Here, ${\cal H}_{{\rm eq},\vec k}(x_{\vec k})$ is the time independent energy of the form~(\ref{FOUR.9}) with
$\omega_{{\rm eq},k} =  (k^2 + r + gS_{\rm eq})$ and
$S_{\rm eq} = \frac{1}{V} \sum_{\vec k} 2\theta_k \sigma_{{\rm eq},\vec k}$ 
is determined self-consistently via the equation 
\be
\sigma_{{\rm eq},\vec k} = \langle x^2_{\vec k} \rangle_{\rm eq} =   \frac{V \beta^{-1}}{2 \theta_k \omega_{{\rm eq},k}}.
\label{Equil.4}
\ee
We can now make the following observations:

\begin{enumerate}

\item   the average energy per mode is given by
\be
\langle {\cal H}_{\vec k} \rangle_{\rm eq}  = \frac{1}{V} \theta_k \omega_{{\rm eq},k} \sigma_{{\rm eq},\vec k} = 
\frac{\beta^{-1}}{2},
\label{Equil.4bis}
\ee
which is the equipartition statement.

\item The correlation length $\xi(T) = [r + gS_{\rm eq}]^{-1/2}$ satisfies
the self-consistency equation
\be
\xi^{-2} = r+ \frac{gT}{V}\sum_{\vec k} \frac{1}{k^2 + \xi^{-2}}.
\label{Equil.5}
\ee
Therefore, defining the critical temperature by the condition 
$\xi^{-1}(T_C) = 0$,
taking the large volume limit and transforming the sum into an integral, we find
\be
T_C = -\frac{r}{g}\frac{(2\pi)^d(d-2)}{\Upsilon_d \Lambda^{d-2}},
\label{Equil.9}
\ee
from which follows that, in order to have a finite critical temperature,
$r$ must be negative and the space dimensionality $d > 2$.

\item Eq.~(\ref{Equil.4}) can be rewritten in the form of the Dyson equation
\be
\sigma_{{\rm eq},\vec k} = \frac{1}{\sigma^{-1}_{0,\vec k} - \Sigma},
\label{Dyson.1}
\ee
where
\be 
\sigma_{0,\vec k} = \frac{V \beta^{-1}}{2\theta_k (k^2 + r)}
\label{Dyson.2}
\ee
is the bare variance, that is the variance of $x_{\vec k}$ in the non interacting Gaussian model,
obtained by setting $g=0$ in Eq.~(\ref{Gin.1}), and  
\be
\Sigma = -\frac{2\theta_k}{V} \beta g S_{\rm eq}
\label{Dyson.3}
\ee
is the tadpole contribution to the self-energy. This is another way to see that the large $N$ limit
corresponds to a mean field approximation of the $N$ vector model.

\end{enumerate}

\subsection{Energy and heat fluctuations}
\label{EnHe}

The fluctuations of energy and heat, during the quench and in the single replica,
are governed by the probability distributions
\be
P(E,t) = \int_{\Omega} d\mathbf{x} \, P[\mathbf{x},t]\delta (E-{\cal H}_{\rm eff}[\mathbf{x},t]),
\label{Init.3}
\ee
\be
P(Q,t,t_w) = \int_{\Omega} d\mathbf{x} d\mathbf{x}_w \, P[\mathbf{x},t;\mathbf{x}_w,t_w] 
\delta (Q-\Delta {\cal H}_{\rm eff}[\mathbf{x},t;\mathbf{x}_w,t_w]),
\label{Init.4}
\ee
where $\Delta {\cal H}_{\rm eff}$ is the difference
\be
\Delta {\cal H}_{\rm eff}[\mathbf{x},t;\mathbf{x}_w,t_w] = {\cal H}_{\rm eff}[\mathbf{x},t] - 
{\cal H}_{\rm eff}[\mathbf{x}_w,t_w].
\label{Init.2}
\ee
We can identify the energy difference with heat because heat fluctuations will be analyzed only 
for the Gaussian model (see Sec.~\ref{para}), in which case
no work is done on or by the system during the quench.

According to the scheme of Sec.~\ref{EH}, $P[\mathbf{x},t]$ and $P[\mathbf{x},t;\mathbf{x}_w,t_w]$
play the role of the prior distributions. Control parameters are the time variables
and the temperatures of the quench. For simplicity, we keep track only of time.
The above integrals cannot be calculated directly, as in the examples of Sec.~\ref{ideal}.
Therefore, 
the calculation of $P(E,t)$ and $P(Q,t,t_w)$ requires the computation of the
corresponding cumulant generating functions, which involves a number of intermediate steps.

First of all, due to the factorization of the initial equilibrium state, which, after
Eq.~(\ref{Equil.1}) is given by
$P_{\rm eq}[\mathbf{x},\beta_I] = \prod_{\vec k} P_{{\rm eq},\vec k}(x_{\vec k},\beta_I)$,
and due to mode independence during the time evolution, the instantaneous
and the joint prior probability densities are also factorized 
\be
P[\mathbf{x},t] = \prod_{\vec k} P_{\vec k}(x_{\vec k},t),
\label{GMD.21}
\ee
\be
P[\mathbf{x},t;\mathbf{x}_w,t_w] =  \prod_{\vec k} P_{\vec k}(x_{\vec k},t,x_{w,\vec k},t_w).
\label{GMD.23}
\ee
The single-mode contributions in the first one are obtained by the integration
\be
P_{\vec k}(x_{\vec k},t) = \int_{-\infty}^{\infty} dx_{\vec k}^{\prime} \, 
R_{\vec k}(x_{\vec k},t|x_{\vec k}^{\prime},0)P_{{\rm eq},\vec k}(x_{\vec k}^{\prime},\beta_I),
\label{Indv.01}
\ee
which yields
\be
P_{\vec k}(x_{\vec k},t)  =  Z^{-1}_{\vec k}(t) e^{-\beta_k(t) {\cal H}_{\vec k}(x_{\vec k},t)},
\label{Indv.1}
\ee
\be
Z_{\vec k}(t)=\sqrt{\frac{\pi V}{\beta_k(t)\theta_k \omega_k(t)}},
\label{Indv.2}
\ee
where $\beta^{-1}_k(t)$ is the effective temperature of the modes with wave vector magnitude $k$, 
defined, as in Eq.~(\ref{Equil.4bis}), from the average energy per degree of freedom~\cite{ciliberto}   
\be
\beta^{-1}_k(t) =   2\langle {\cal H}_{\vec k}(t) \rangle = \frac{2}{V} \theta_k \omega_k(t)\sigma_k(t),
\label{Indv.2bis}
\ee
with
\be
\sigma_k(t) = \langle x_{\vec k}^2(t) \rangle,
\label{Indv.2tris}
\ee
from which, using Eq.~(\ref {FOUR.70}) at equal times, follows  
\be
\beta^{-1}_k(t) 
=  \frac{2}{V} \theta_k \omega_k(t)
[G^2_k(t,0) \langle x^2_{\vec k}(0) \rangle + 2 T_{F,k} \int_0^t dt^{\prime} \, G^2_k(t,t^{\prime})]. 
\label{Indv.2tris}
\ee
As anticipated in the Introduction, it is evident from the above expression that 
after the quench this quantity acquires a $k$-dependence, signaling the breaking of
equipartion and departure from equilibrium. Interestingly, 
$k$-dependent effective temperatures were previously introduced by Padilla and Ritort~\cite{Padilla}
in the study of the relaxational dynamics of a glassy system.

Similarly, using Eqs.~(\ref{FOUR.4}) and~(\ref{Indv.1}), the single-mode joint probabilities are given by
\begin{eqnarray}
P_{\vec k}(x_{\vec k},t,x_{w,\vec k},t_w) & = &  R_{\vec k}(x_{\vec k},t|x_{w,\vec k},t_w)P_{\vec k}(x_{w,\vec k},t_w) \nonumber \\
& = & Z^{-1}_{\vec k}(t,t_w) \exp \left \{-\frac{1}{2} U_{\vec k}(x_{\vec k},t,x_{w,\vec k},t_w) \right \},
\label{Indv.3}
\end{eqnarray}
with
\begin{eqnarray}
& & U_{\vec k}(x_{\vec k},t,x_{w,\vec k},t_w) = \nonumber \\
& & \frac{1}{[1-\rho^2_k(t,t_w)]} \left \{ \frac{x^2_{\vec k}}{\sigma_k(t)}
-\frac{2\rho_k(t,t_w)}{\sqrt{\sigma_k(t)\sigma_k(t_w)}}x_{\vec k}x_{w,\vec k} + \frac{x^2_{w,\vec k}}{\sigma_k(t_w)} \right \},
\label{Indv.4}
\end{eqnarray}
\be
\rho_k(t,t_w) = G_k(t,t_w)\sqrt{\frac{\sigma_k(t_w)}{\sigma_k(t)}},
\label{Indv.5}
\ee
\be
Z_{\vec k}(t,t_w) = 2\pi \sqrt{[1-\rho^2_k(t,t_w)]\sigma_k(t)\sigma_k(t_w)}.
\label{Indv.6}
\ee
The above factorizations imply, in turn, the factorization of the 
corresponding moment generating functions
\be
K_{{\cal H}}(z,t) =   \prod_{\vec k}K_{{\cal H},\vec k}(z,t),
\label{EnHe.1}
\ee
\be
K_{{\Delta \cal H}}(z,t,t_w) = \prod_{\vec k} K_{{\Delta \cal H},\vec k}(z,t,t_w),
\label{EnHe.2}
\ee
whose single-mode factors are given by
\be
K_{{\cal H},\vec k}(z,t) =  \int_{-\infty}^{\infty} dx_{\vec k} \, P_{\vec k}(x_{\vec k},t)
e^{z{\cal H}_{\vec k}(x_{\vec k},t)} =  \frac{1}{\sqrt{1 - \beta^{-1}_k(t)z}}
\label{EnHe.5}
\ee
and
\begin{eqnarray}
& & K_{{\Delta \cal H},\vec k}(z,t,t_w)  = \nonumber \\
& & Z^{-1}_{\vec k}(t,t_w) \int_{-\infty}^{\infty} dx_{\vec k} dx_{w,\vec k} \, 
e^{-\frac{1}{2} U_{\vec k}(x_{\vec k},t,x_{w,\vec k},t_w) + z[{\cal H}_{\vec k}(x_{\vec k},t) - {\cal H}_{\vec k}(x_{w,\vec k},t_w)] } = \nonumber \\
& & \frac{1}{\sqrt{[1 - {\cal Q}_{-,k}(t,t_w)z][1 - {\cal Q}_{+,k}(t,t_w)z]}}. 
\label{EnHe.6}
\end{eqnarray}
Omitting time arguments, the quantities appearing above are defined by
\be
{\cal Q}_{-,k} = - \left [ \frac{a_k}{2} + \sqrt{\left (\frac{a_k}{2} \right )^2 + b_k} \right ],
\label{EnHe.7}
\ee
\be
{\cal Q}_{+,k} =  \left [ \sqrt{\left (\frac{a_k}{2} \right )^2 + b_k} \right ] - \frac{a_k}{2},
\label{EnHe.8}
\ee
with
\be
a_k = \beta^{-1}_k(t_w) - \beta^{-1}_k(t),
\label{EnHe.9}
\ee
\be
b_k = [1-\rho_k^2(t,t_w)] \beta^{-1}_k(t_w)\beta^{-1}_k(t).
\label{EnHe.10}
\ee
In the following, for simplicity, heat fluctuations will be considered only in the case of
a zero temperature quench $(T_F=0)$. In this case, using definitions, it is easy to show that $b_k=0$ and
\be
{\cal Q}_{-,k} = \left [ \beta^{-1}_k(t) - \beta^{-1}_k(t_w) \right ], \;\;\; {\cal Q}_{+,k} = 0.
\label{EnHe.100}
\ee 
Hence, dropping ${\cal Q}_{+,k}$, replacing ${\cal Q}_{-,k}$ with ${\cal Q}_k$, 
and using the general result
\be
\frac{1}{2\pi i} \int_{\alpha -i \infty}^{\alpha +i \infty} dz \, \frac{e^{-zx}}{\sqrt{1-\kappa z}} 
= \frac{e^{-\frac{x}{\kappa}}}{\sqrt{ \pi \kappa x}} \theta(\kappa x),
\label{EI.3}
\ee
where $\theta$ is, again, the Heaviside step function, from Eqs.~(\ref{EnHe.5}) and~(\ref{EnHe.6})
we find the probabilities of energy and heat fluctuations in the $\vec k$-mode
\be 
P_{\vec k}(E_k) = \frac{e^{-\beta_kE_k}}{\sqrt{ \pi \beta^{-1}_k E_k}} \theta(\beta^{-1}_k E_k),
\label{hr.10}
\ee
\be 
P_{\vec k}(Q_k) = \frac{e^{-\frac{Q_k}{{\cal Q}_k}}}{\sqrt{ \pi {\cal Q}_k Q_k}} \theta({\cal Q}_k Q_k),
\label{hr.1}
\ee
showing that, as $\beta^{-1}_k/2$ is the average energy per mode, so ${\cal Q}_k/2$ is the average heat exchanged
per mode.

The square roots appearing in Eqs.~(\ref{EnHe.5}) and~(\ref{EnHe.6}) impose restrictions on the
values that $z$ can take. As we shall see, the spectrum $[\beta_k]$ 
can contain a positive as well as a negative branch. Denoting by $\beta_{\rm max}$ the upper edge of the negative branch
and by $\beta_{\rm min}$ the lower edge of the positive branch,
the reality of $K_{{\cal H}}(z,t)$ restricts 
the range of allowed $z$ values to the interval
\be
[\beta_{\rm max},\beta_{\rm min}].
\label{R.10}
\ee
Similarly, the domain of definition of $K_{{\Delta \cal H}}(z,t,t_w)$ is
\be
[{\cal Q}^{-1}_{\rm max},{\cal Q}^{-1}_{\rm min}] ,
\label{R.11}
\ee
where ${\cal Q}^{-1}_{\rm max}$ and ${\cal Q}^{-1}_{\rm min}$ are the upper and the lower edges of the
negative and positive branches, respectively, of the spectrum of inverse average heat.

Using the above results, the
cumulant generating functions~(\ref{gen.7bis}) are given by
\be
\lambda_{{\cal H}}(z,t)  =  - \frac{1}{2V} \sum_{\vec k}\ln [1 - \beta_k^{-1}(t)z] 
\label{EnHe.3}
\ee
and
\be
\lambda_{{\Delta \cal H}}(z,t,t_w)   
=  - \frac{1}{2V} \sum_{\vec k}\left [\ln (1 - {\cal Q}_{k}(t,t_w)z) \right ].
\label{EnHe.4}
\ee
Then, in order to complete the calculation of the probabilities~(\ref{Init.3}) and~(\ref{Init.4}), 
the rate functions $I_{\cal H}(e,t)$ and $I_{\Delta {\cal H}}(q,t,t_w)$ must be computed. This,
according to Eq.~(\ref{gen.9}), requires
the solution of the saddle point equations
\be
e = \widetilde{F}_{{\cal H}}(z,t,V),
\label{MDF.4}
\ee
\be
q = \widetilde{F}_{\Delta {\cal H}}(z,t,t_w,V),
\label{MDF.5}
\ee
where $e=E/V$, $q=Q/V$ are the densities. The functions in the right hand sides, recalling Eq.~(\ref{GELLIS.01}),
are given by
\be
\widetilde{F}_{{\cal H}}(z,t,V) = \frac{1}{V} \sum_{\vec k} \langle {\cal H}_{\vec k} \rangle_z,
\label{MDF.6}
\ee
\be
\widetilde{F}_{\Delta {\cal H}}(z,t,t_w,V) = \frac{1}{V} \sum_{\vec k} \langle {\Delta \cal H}_{\vec k} \rangle_z,
\label{MDF.7}
\ee
where
\be
\langle {\cal H}_{\vec k} \rangle_z = \frac{1}{2[\beta_k(t) - z]},
\label{MDF.6bis}
\ee
\be
\langle \Delta {\cal H}_{\vec k} \rangle_z = \frac{1}{2[{\cal Q}^{-1}_{k} - z]},
\label{MDF.7bis}
\ee
are the average energy and heat per mode, in the corresponding biased ensembles. 
The formal solutions can be written as
\be
z^*(e,t) = \widetilde{F}_{{\cal H}}^{-1}(e,t,V),
\label{SPMF.3}
\ee
\be
z^*(q,t,t_w) = \widetilde{F}_{\Delta {\cal H}}^{-1}(q,t,,t_w,V),
\label{SPMF.3}
\ee
where $\widetilde{F}_{{\cal H}}^{-1}$ and $\widetilde{F}_{\Delta {\cal H}}^{-1}$ 
are the inverse, with respect to $z$, of the functions
defined by Eqs.~(\ref{MDF.6}) and~(\ref{MDF.7}).
In order to discuss the actual existence of these solutions,
we shall first consider, in the next section, the simpler case of the quench 
within the paramagnetic phase and then,
in the subsequent section, the phase-ordering process in the quench
from above to below $T_C$.

Before concluding this section, let us make the following observation. Recalling Eq.~(\ref{Indv.2bis}),
we may rewrite Eq.~(\ref{MDF.6bis}) as 
\be
\langle {\cal H}_{\vec k} \rangle_z = \frac{1}{\langle {\cal H}_{\vec k}(t) \rangle^{-1} -  2z},
\label{Dyson.5}
\ee
in which the biased and the prior averages of the energy per mode enter in the same formal
relationship as the dressed and the bare average in the Dyson equation~(\ref{Dyson.1}), 
with $2z$ playing the role of the tadpole self-energy. Therefore, biased expectations can
be viewed as arising from the mean field approximation on an underlying interacting theory,
whose free limit is given by the prior expectations. This will turn out to be essential for
the distinction between condensation as a typical phenomenon or as a rare fluctuation.
Clearly, the same considerations apply to Eq.~(\ref{MDF.7bis}), where ${\cal Q}_{k}$
and $\langle \Delta {\cal H}_{\vec k} \rangle_z$ are the bare and the dressed heat exchanged,
respectively.

\section{Quench of the paramagnet}
\label{para}

\begin{figure}
\centering
\begin{tabular}{cc}
\epsfig{file=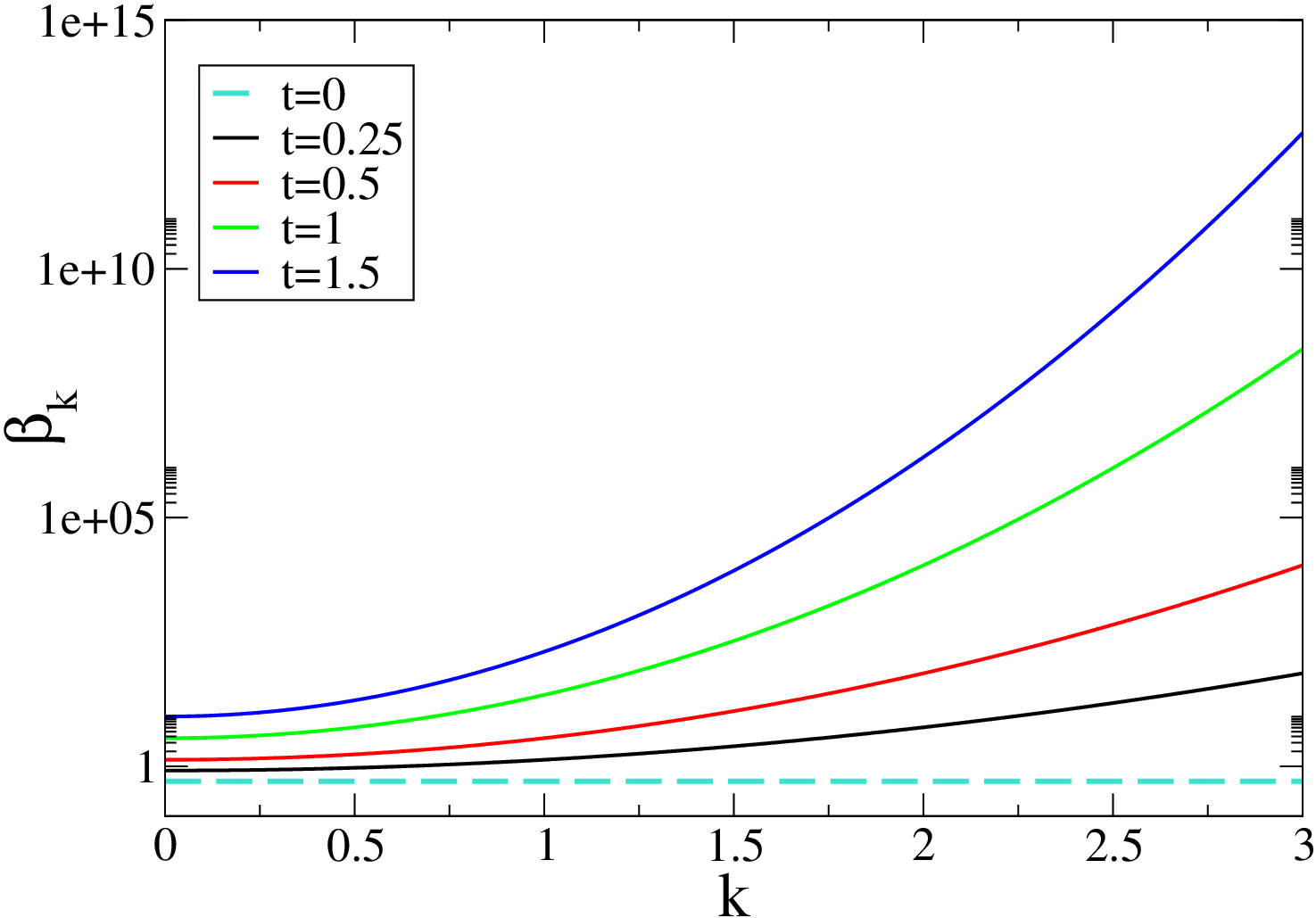,width=0.5\linewidth,clip} &
\epsfig{file=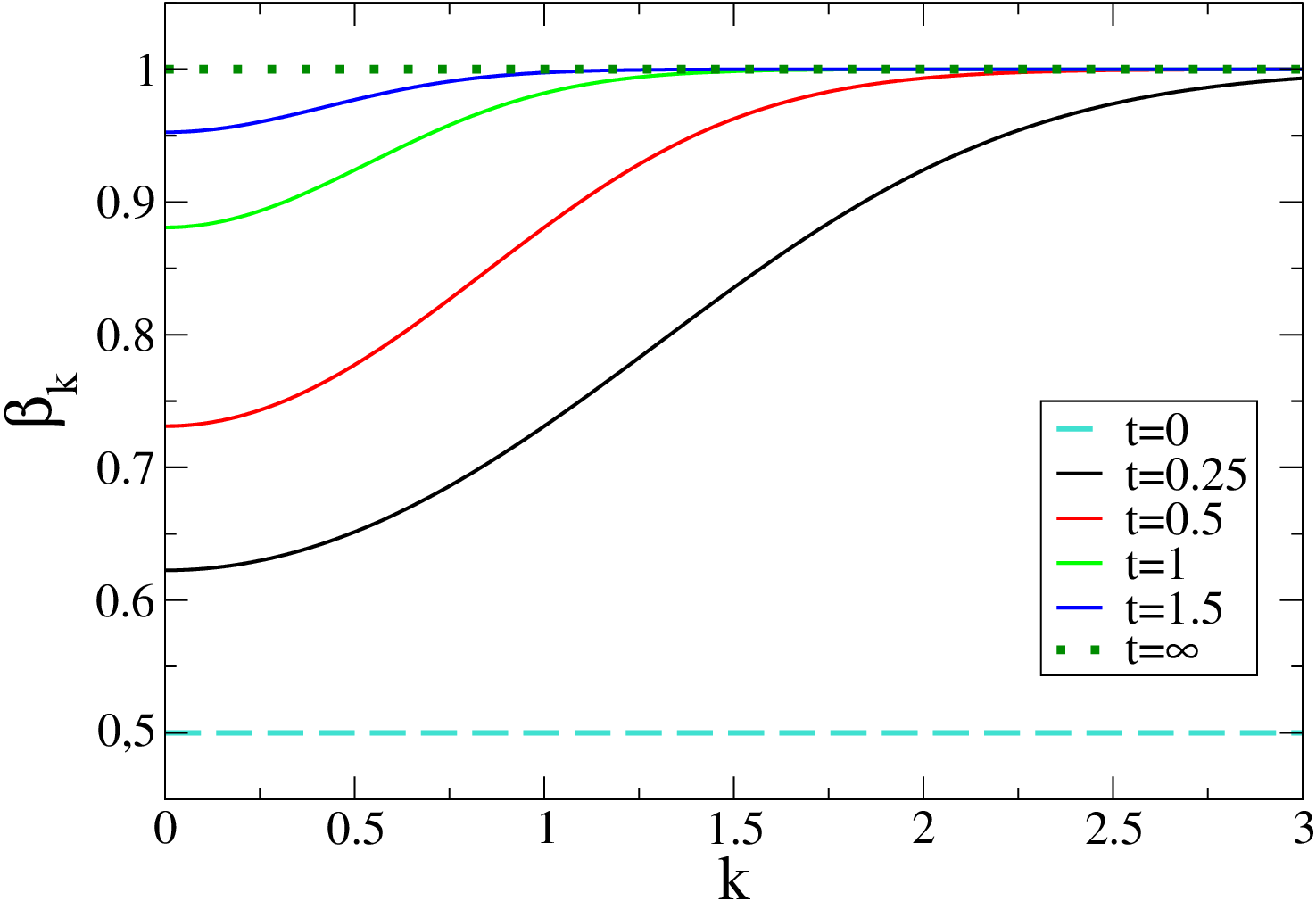,width=0.5\linewidth,clip}
\end{tabular}
\caption{Time evolution of the inverse temperature spectrum with 
$T_I=2$,  $r=1$, $T_F=0$ (left panel), $T_F=1$ (right panel).}
\label{beta_k}
\end{figure}

If the quench is limited to the the paramagnetic phase, the problem can be simplified
by taking $r > 0$ and by dropping the non linear $g$ term in Eq.~(\ref{Gin.5}), which would
only modify inessential quantitative details. Hence we shall work with the Gaussian model
\be
{\cal H}[\varphi] = \frac{1}{2} \int_V d \vec x \, [(\nabla \varphi)^2 + r \varphi^2 (\vec x)],
\label{GMD.1}
\ee
which is the basic non interacting model in the
theory of phase transitions~\cite{Goldenfeld}. With $r > 0$ there is no transition
and the system is paramagnetic at any temperature. In spite of the apparent triviality
of the model, 
when fluctuations of macrovariables are considered non trivial behavior may arise,
as we shall see shortly. 
With this choice of parameters the dispersion relation~(\ref{Gin.6})
becomes time independent $\omega_k= k^2 + r$. The effective temperature~(\ref{Indv.2bis})
takes the simple form 
\be
\beta^{-1}_k(t) = \Delta T e^{-2\omega_k t} + T_F,  
\label{eff.1}
\ee
where $\Delta T= T_I - T_F$ is the temperature jump across the quench.
The behavior of $\beta_k(t)$ is displayed in Fig.~\ref{beta_k}. In the left panel for a quench to $T_F=0$
and in the right panel for a quench to the finite final temperature $T_F > 0$. 
Initially, equipartition holds and the spectrum is flat with $\beta_k(t=0) = \beta_I$,
Then, as the system is put off equilibrium and relaxation begins, 
the temperature of the different modes spread out, signaling 
the breakdown of equipartition. 
The $k=0$ mode is the hottest, while the temperature decreases as $k$ increases. This is due to the
fact that the relaxation time of the different modes is $k$-dependent and decreases as $k$ increases.
Eventually all $\beta_k(t)$ relax to the same final value $\beta_F$, as the system
equilibrates again and equipartition is restored.

\subsection{Energy fluctuations}
\label{largevolume}

From Eq.~(\ref{eff.1}) follows that in the temperature spectrum there exists only the positive branch,
with the lower edge at $\vec k =0$, namely $\beta_{\rm min} = \beta_0$ (see Fig.~\ref{beta_k}). Therefore, 
$\widetilde{F}_{{\cal H}}(z,t,V)$ is well defined for $z \leq \beta_0$.
In order to ease notation, from now on we shall omit the explicit time dependence of $\beta_{\vec k}(t)$.
As long as $V$ is finite, $\widetilde{F}_{{\cal H}}(z,t,V)$  increases monotonically
from $0$ to $\infty$ as $z$ varies from $-\infty$ up to  
$\beta_0$. Consequently $z^*(e,t)$, given by Eq.~(\ref {SPMF.3}), exists for any non negative $e$. 
However, as $V$ becomes large, $\widetilde{F}_{{\cal H}}(z,t,V)$
may become singular, depending on the strength of the divergence at the lower
edge of the spectrum.

Let us, then, proceed as we have done for the Bose gas in Sec.~\ref{ideal},
by separating the divergent term
from the sum and transforming the rest of it into an integral
\be
e = \frac{1}{V} \langle {\cal H}_0 \rangle_{z} + F_{{\cal H}}(z,t),
\label{condav.1}
\ee 
where $\langle {\cal H}_0 \rangle_{z}$ is the biased expectation of the zero mode energy,
defined by Eq.~(\ref{MDF.6bis}), and
\be
F_{{\cal H}}(z,t) = \frac{\Upsilon_d}{2} \int_0^{\Lambda} d k \, \frac{k^{d-1}}{\beta_{k} -z}
\label{condav.20}
\ee
is the contribution arising from all the other modes. Then, as in the case of the Bose gas,
the focus is shifted on the behavior of 
$F_{{\cal H}}(z,t)$ as $z$ approaches the lower edge of the spectrum $\beta_0$. 
Keeping $t$ fixed and denoting by 
\be
e_C(t) = F_{{\cal H}}(\beta_0,t)
\label{condav.3}
\ee
the upper bound on $F_{{\cal H}}(z,t)$, we have that $e_C(t)$ diverges for $d \leq 2$,
while it is finite for $d > 2$, because the denominator under the integral 
vanishes like $k^2$ for small $k$.
Conversely, for $d > 2$  the singularity is integrable and $e_C(t)$ is finite.
Hence, for $d > 2$ we have
\be
\frac{1}{V} \langle {\cal H}_0 \rangle_{z^*}  = \left \{ \begin{array}{ll}
        {\cal O}(1/V) ,\;\; $for$ \;\; e \leq e_C(t),\\
        e-e_C(t) ,\;\; $for$ \;\; e > e_C(t),
        \end{array}
        \right .
        \label{condav.4}
        \ee
and 
\be
z^*(e,t)  = \left \{ \begin{array}{ll}
        F_{{\cal H}}^{-1}(e,t) < \beta_0  ,\;\; $for$ \;\; e \leq e_C(t),\\
        \beta_0 ,\;\; $for$ \;\; e > e_C(t).
        \end{array}
        \right .
        \label{condav.5}
        \ee

\subsubsection{Condensation on average}

The statement, in Eq.~(\ref{condav.4}), is that the expectation value
of the zero mode energy, in the biased ensemble, makes the transition from microscopic
to macroscopic as $e$, which from Eq.~(\ref{GELLIS.01}) is the average value of the
total energy, crosses the critical value $e_C(t)$. 
Hence, this is a condensation transition showing up as an average feature, 
driven by $e$ and at a fixed time $t$ after the quench. 
Though the mechanism of the transition is entirely analogous to BEC, the important
difference is that this is exclusively an out of equilibrium phenomenon,
which cannot take place in equilibrium as now will be explained. 

\begin{figure}[h]

	\vspace{1cm}

    \centering
   \rotatebox{0}{\resizebox{.4\textwidth}{!}{\includegraphics{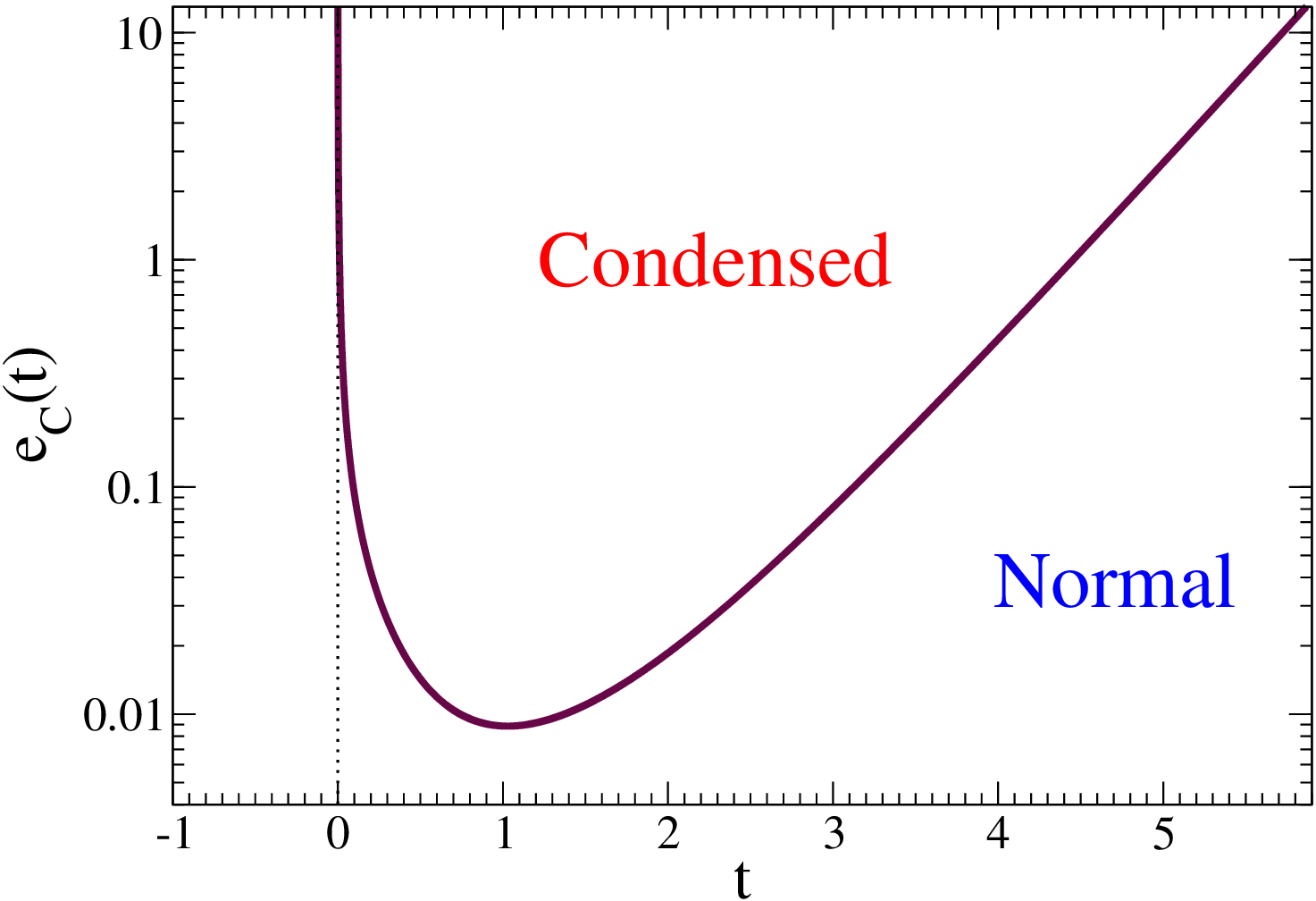}}}
\hspace{1cm}
   \rotatebox{0}{\resizebox{.4\textwidth}{!}{\includegraphics{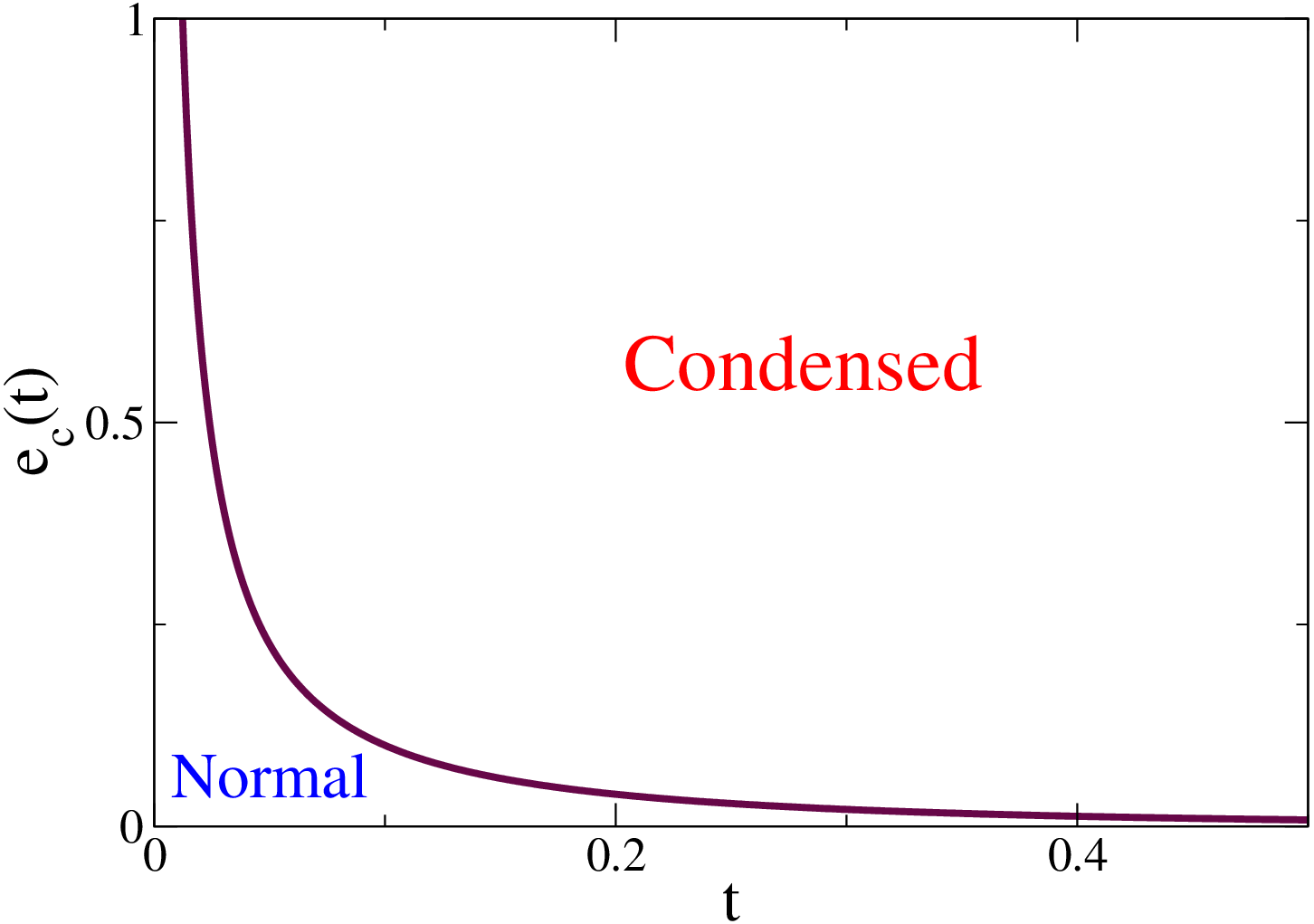}}}
   \caption{Energy phase diagram in the paramagnetic quench, with $d=3, r=1,T_I=1$. 
Left panel $T_F=0.2$, right panel $T_F=0$.}
\label{ph_diag_S}
\end{figure}

If $t$ is let to vary, $e_C(t)$ moves along the critical line on the $(t,e)$ plane, 
which separates the condensed phase (above) from the normal phase (below), as
depicted in Fig.~\ref{ph_diag_S} for $T_F > 0$ and
$T_F=0$. In both cases the curves diverge to infinity at $t=0$. This is
due to the fact that there cannot be a condensed phase
in the initial equilibrium state,
since the denominator under the integral~(\ref{condav.20})
vanishes identically for all $k$, producing the divergence of $e_C(T_I)$ for any space dimensionality.
However, as soon as the system is put off equilibrium by the quench,
a number of new features appear: equipartition is broken with 
the spectrum of inverse temperatures developing a minimum at 
$k=0$, which causes the convergence of the integral defining $e_C(t)$. 
Therefore, after the quench, $e_C(t)$ 
drops down from infinity. If $T_F > 0$, the critical curve reaches a minimum
and then rises again toward infinity as the system equilibrates to the final finite temperature.
Instead, if $T_F=0$, the threshold $e_C(t)$ keeps on decreasing and eventually vanishes, since
as time goes on the modes freeze starting from the higher $k$'s and it is possible to keep
a finite amount of energy in the system
only if a finite fraction of it is contributed by the zero mode, which is 
the slowest to freeze. The phase diagram in the right panel of Fig.~\ref{ph_diag_S},
extending to arbitrary positive energies, although $T_F=0$, is to be understood in the framework of
the biased ensemble, where, according to the remarks made about Eq.~(\ref{Dyson.5}) at the
end of Sec.~\ref{EnHe}, the mode temperatures are renormalized by the bias. In particular, 
the renormalized temperature of the zero mode
can become arbitrarily high by taking $z$ sufficiently close to $\beta_0$. 

Finally, notice that
the non monotonic shape of the critical line, when $T_F > 0$, implies that the
the transition driven by $t$ is re-entrant when the total energy $e$ is kept fixed
to a value above the minimum of the critical line.

\subsubsection{Condensation of fluctuations}
\label{condfluct}

Let us now comment the phase diagram in the context of the prior ensemble. 
Most of the results in this subsection have been derived in Ref.~\cite{PREZ}.
Recalling that the effective temperature represents the average
energy per mode, it is clear from Fig.~\ref{beta_k} that in the prior
ensemble no singularity appears in the average properties.
In other words, the prior ensemble does not produce any phase
transition on average.
However, above we have seen that to the singular behavior
of $z^*$, in Eq.~(\ref{condav.5}), there corresponds a transition in the biased
ensemble which, as mentioned, is analogous to BEC in the grand canonical ensemble.
In order to explore the counterpart of this transition in the behavior of
the fluctuations, let us go back to Eq.~(\ref{Init.3}). 
Denoting by $\{E_{\vec k}\}$ a microscopic energy configuration, we may write
\be
P(E,t) = \int \prod_{\vec k} dE_{\vec k} \,
P(\{E_{\vec k}\},t) \delta (E- \sum_{\vec k} E_{\vec k}),
\label{condfluct.1}
\ee
where the probability of the configuration is given by
\be
P(\{E_{\vec k}\},t) =  \prod_{\vec k} P_{\vec k}(E_{\vec k},t)
\label{condfluct.2}
\ee
and $P_{\vec k}(E_{\vec k},t)$ has been computed in Eq.~(\ref{hr.10}).
The statement in Eq.~(\ref{condfluct.1}) is simply that, once $E$ has been given,
the allowed microscopic events $\{E_{\vec k}\}$ are those on the hypersurface defined by the constraint
$E = \sum_{\vec k} E_{\vec k}$ and that the probability $P(E,t)$ is obtained by summing
over this energy shell.
On the other hand, $P(E,t)$ is also given by
\be
P(E,t) \sim e^{-VI_{\cal H}(e,t)},
\label{condfluct.5}
\ee
where
\be
I_{\cal H}(e,t) = z^*(e,t)e + \lambda_{\cal H}(z^*,t).
\label{condfluct.6}
\ee
Taking $e > e_C(t)$ and recalling that in this case $z^*$ sticks~\cite{Berlin} to $\beta_0$, we may rewrite  
\be
I_{\cal H}(e,t) = \beta_0(e-e_C) + I_{\cal H}(e_C,t),
\label{condfluct.7}
\ee
from which follows
\be
P(E,t) \sim e^{-\beta_0V(e-e_C(t))}P(E_C,t).
\label{condfluct.8}
\ee
Hence, keeping into account the result~(\ref{hr.10}), in place of Eq.~(\ref{condfluct.1}) we have
\begin{eqnarray}
P(E,t) & = & \int dE_{0} P_0(E_{0},t)\delta (E_0- (E-E_C)) \nonumber \\
& \times & \int \prod_{\vec k \neq 0} dE_{\vec k} \,
P_{\vec k}(E_{\vec k},t) \delta (E_C- \sum_{\vec k \neq 0} E_{\vec k}),
\label{condfluct.10}
\end{eqnarray}
which means that, for $e > e_C(t)$, the probability of the configurations $\{ E_{\vec k}\}$
is concentrated on the subset of the energy shell singled out by 
the additional condition $E_0 = E - E_C$.
This is condensation of fluctuations, in the sense that an energy fluctuation above threshold
can occur only if the macroscopic fraction $E-E_C$ of it is contributed by the zero mode.
As anticipated in Sec.~\ref{EH}, the remarkable feature of this transition is that it
takes place in a non interacting system, like the Gaussian model, in which
no transition on average can take place, in and out of equilibrium. The explanation is in Eq.~(\ref{Dyson.5}),
which shows how the bias generates the interaction sustaining the transition, and the bias is
generated once the size of the fluctuation has been fixed.

\subsection{Heat fluctuations}

The study of heat fluctuations proceeds along the same lines, keeping
into account, however, that now there are two times $t$ and $t_w$ to keep track of.
Recalling that for heat we consider only the quench to $T_F=0$,
from Eqs.~(\ref{eff.1}) and~(\ref{EnHe.100}) follows that in the spectrum
of inverse average heat there is only the negative branch with upper edge at
${\cal Q}^{-1}_{{\rm max}}$. Consequently, the domain of definition of
$\widetilde{F}_{\Delta {\cal H}}(z,t,t_w,V)$ extends to all
$z \geq {\cal Q}^{-1}_{{\rm max}}$. 
Defining by $k_{\rm max}$ the magnitude of the wave vector at the edge of the spectrum,
that is ${\cal Q}^{-1}_{k_{\rm max}} = {\cal Q}^{-1}_{{\rm max}}$, and proceeding as in Eq.~(\ref{condav.1}) by 
separating the most divergent term
plus transforming the rest of the sum into an integral, Eq.~(\ref{MDF.5}) becomes
\be
q = \frac{1}{2V} \langle \Delta {\cal H}_{\vec {k}_{\rm max}} \rangle_z    +  F_{\Delta {\cal H}}(z,t,t_w),
\label{TZ.5}
\ee
with
\be
F_{\Delta {\cal H}}(z,t,t_w) = \frac{\Upsilon_d}{2}\int_0^{\Lambda} d k \, \frac{k^{d-1}}{{\cal Q}^{-1}_{k} -z}.
\label{condav.2}
\ee
This is a negative function, monotonically increasing from the lower bound 
\be
q_C(t,t_w) = F_{\Delta {\cal H}}({\cal Q}^{-1}_{k_{\rm max}},t,t_w)
\label{TZ.7}
\ee
toward zero as $z$ varies from ${\cal Q}^{-1}_{k_{\rm max}}$ to $\infty$. Again, the crucial question is
whether this bound is finite or infinite
and this depends on whether $k_{\rm max} = 0$, or $k_{\rm max} > 0$. In fact, if
$k_{\rm max} > 0$ the singularity is not integrable and 
$q_C(t,t_w)$ diverges negatively. Conversely, if $k_{\rm max} = 0$  
the denominator vanishes like $k^2$ for small $k$ and, as
in the case of energy, $q_C(t,t_w)$ is finite for $d > 2$.
If this is the case, there is condensation of heat fluctuations in
the $k=0$ mode when $q < q_C(t,t_w)$.

\begin{figure}[h]

	\vspace{1cm}

    \centering
   \rotatebox{0}{\resizebox{.4\textwidth}{!}{\includegraphics{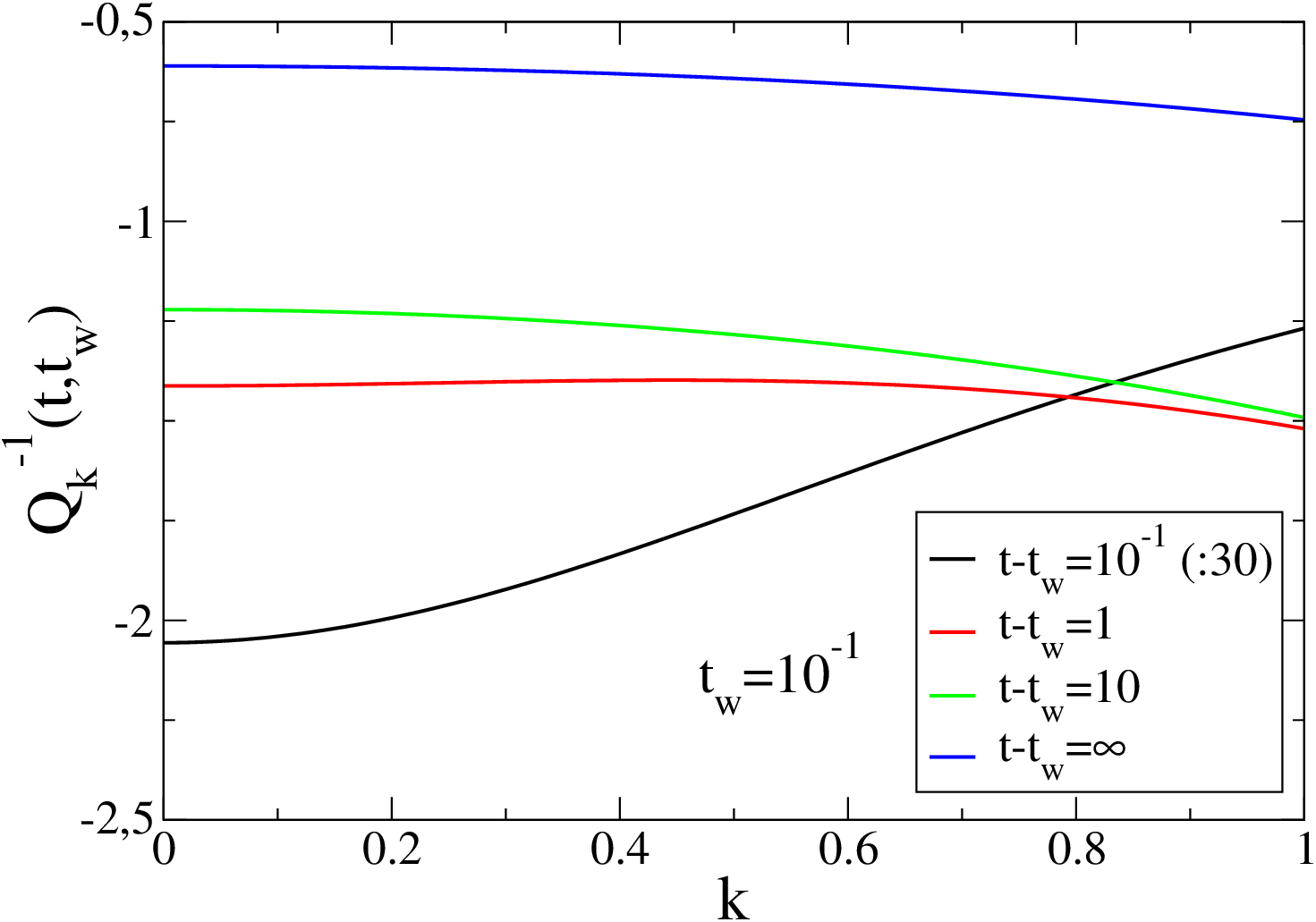}}}
\hspace{1cm}
   \rotatebox{0}{\resizebox{.4\textwidth}{!}{\includegraphics{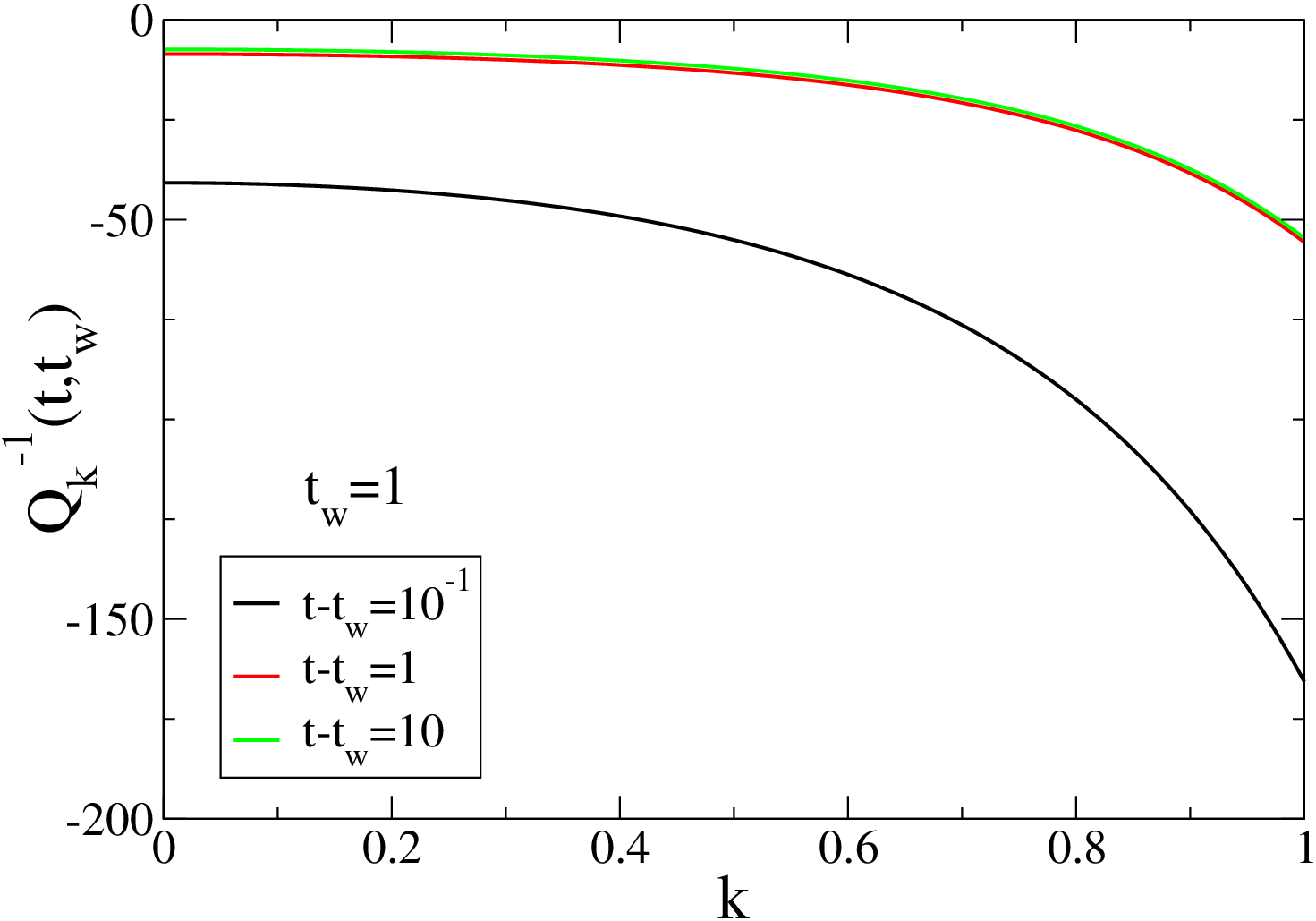}}}
\caption{Time evolution of the inverse heat spectrum in the Gaussian model with $T_I=1$, $r=1$ and
$T_F=0$. Left panel $t_w = 0.1$, right panel $t_w = 1$.}
\label{Gaussian1}
\end{figure}

Thus, in order to establish the occurrence of condensation, it is necessary
to find $k_{\rm max}$. 
For $t_w < \tau/2$, where
$\tau = 1/r$ is the largest relaxation time, and for $t < \widetilde{t}$, where
$\widetilde{t}$ is defined by
\be
\frac{1}{2(\widetilde{t} - t_w)} \ln (\widetilde{t}/t_w) = r,
\label{TZ.8}
\ee
we have
\be
k^2_{\rm max} = \frac{1}{2(t - t_w)} \ln (t/t_w) - r > 0,
\label{TZ.9}
\ee
which implies that $k_{\rm max} \rightarrow 0$ as $t \rightarrow \widetilde{t}$.
Introducing the characteristic
length $\ell(t,t_w)$ defined by
\be
\ell^{-2}(t,t_w) \equiv \frac{1}{2(t - t_w)} \ln (t/t_w)
\label{chrl.1}
\ee
and recalling that $\xi = r^{-1/2}$ is the equilibrium correlation length, 
$\widetilde{t}$ is the time at which $\ell(\widetilde{t},t_w) = \xi$ and Eq.~(\ref{TZ.9}) can be rewritten as
\be
k^2_{\rm max} = \ell^{-2}(t,_wt) - \xi^{-2}.
\label{TZ.9bis}
\ee
Therefore, in order to have $k_{\rm max} > 0$, the system must be out of equilibrium, both
because $t_w < \tau$ and because $\ell(t,t_w) < \xi$ (see left panel of Fig.\ref{Gaussian1}). Namely, there are conditions
on both the waiting time $t_w$ and the time difference $t-t_w$.
Conversely, if either one or both of these conditions are violated, i.e. if $\ell(t,t_w) \geq \xi$
and/or $t_w \geq \tau$, then  $k_{\rm max} = 0$ (see right panel of Fig.\ref{Gaussian1}).

\begin{figure}
\begin{center}
\epsfysize=6.0cm \epsffile{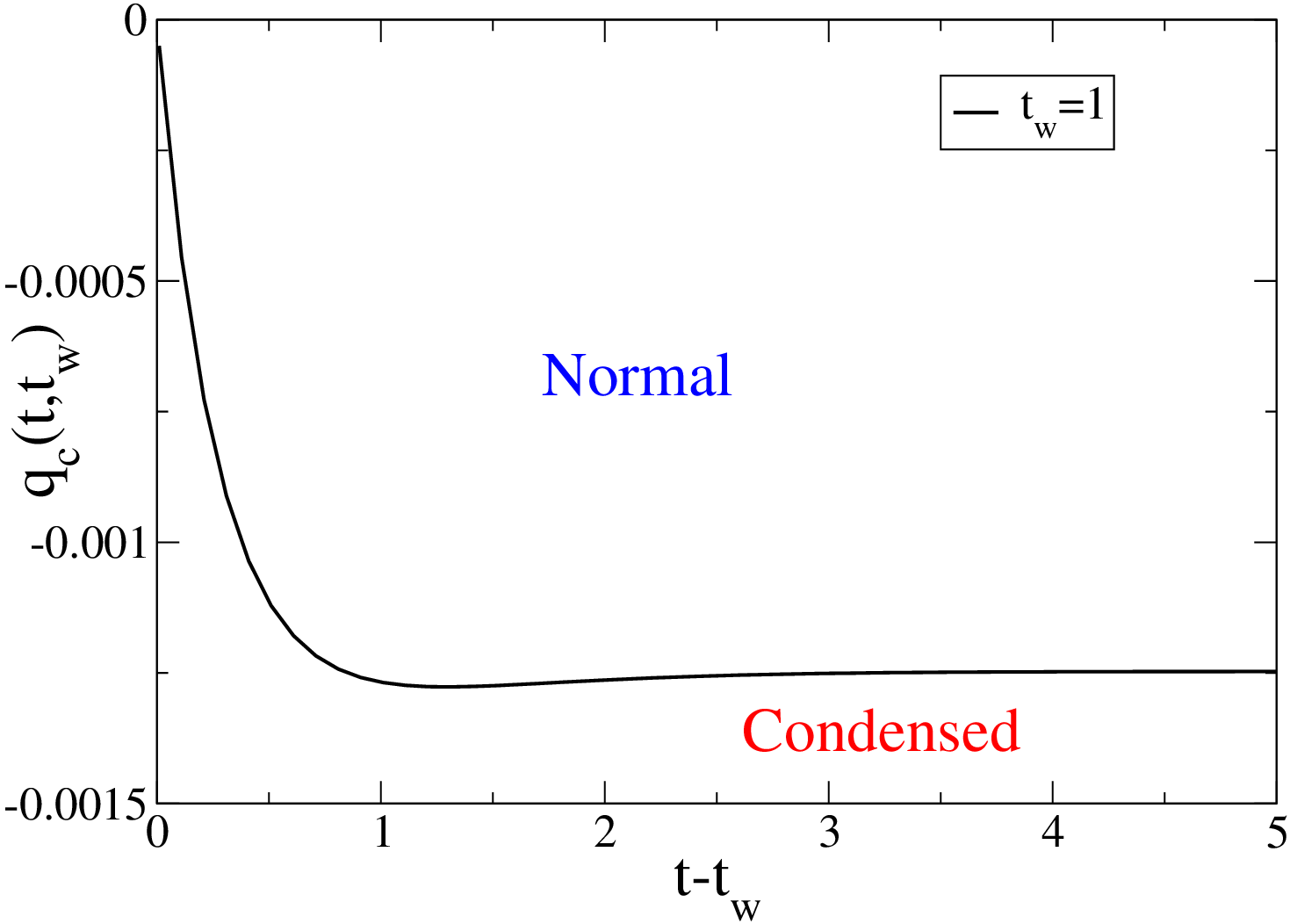} 
\caption{Heat phase diagram for the Gaussian model with $T_I=1$, $r=1$ and
$T_F=0$.}
\label{Gaussian4}
\end{center}
\end{figure}

The phase diagram, obtained by plotting $q_C(t,t_w)$ for fixed $t_w=\tau$ as a function of $t-t_w$,
is shown in Fig.~\ref{Gaussian4}.

\section{Quench of the ferromagnet}
\label{ferro}

In the preceding section we have seen that, even in the simplified context of the 
Gaussian model, fluctuations of energy and heat may exhibit non trivial behavior,
highlighted by the occurrence of condensation transitions. 
The interesting question is if, and to what extent, the picture is modified when
the system gets trapped out of equilibrium
by the phase-ordering process following the quench to below $T_C$. As we shall see, 
in that case there arise important qualitative differences
in the behavior of the fluctuations, which turn out to be strong indicators
that the system does not equilibrate. 

We shall now consider the full large $N$ model with $r < 0$ and,
for simplicity, $T_F=0$  with $T_I$ 
well above $T_C$. Precisely, we shall take $T_I$ such that
the correlation length is of the order of the shortest meaningful length
scale, that is $\xi(T_I) = \Lambda^{-1}$.  For $d=3$, from Eq.~(\ref{Equil.5}) follows
\be
\xi^{-2}= \frac{g \Lambda}{2 \pi^2}(T-T_C) -\frac{g T}{2 \pi^2}\xi^{-1} \arctan (\Lambda \xi).
\label{initl.1}
\ee
Hence, taking $\Lambda=g=-r=1$, $T_C=2\pi^2=19.7$ and imposing $\xi=1$, one gets $T_I=(4\pi)^2/(4-\pi)=183.9$. 
We shall adopt these numerical values of the parameters in the numerical calculations and
shall we limit the discussion to energy fluctuations, since heat has been studied in
detail in Ref.~\cite{prec}.

\subsection{Effective temperature spectrum}

\begin{figure}
\begin{center}
\epsfysize=6.0cm \epsffile{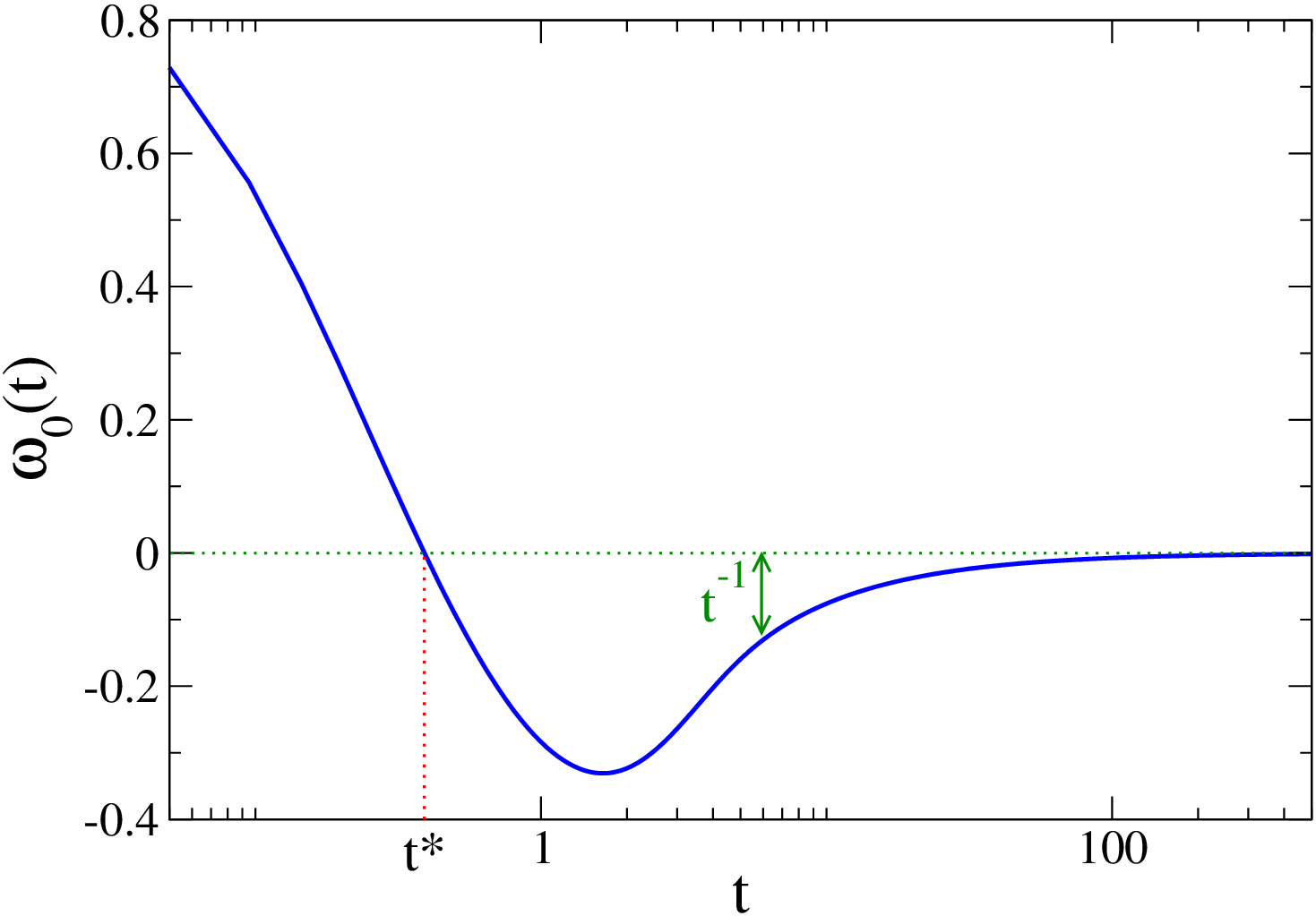} \caption{Time dependence of $\omega_0(t)$
in the large $N$ model quenched to $T_F=0$.} 
\label{omegazero}
\end{center}
\end{figure}

The first step in the study of energy fluctuations
is the understanding of the pattern of equipartition breaking, encoded
into the effective temperature spectrum by Eq.~(\ref{Indv.2bis}).  
The key quantity is $\omega_0(t)$, whose time evolution, depicted in
Fig.~\ref{omegazero}, has been obtained solving numerically Eq.~(\ref{Gin.12}) for $G_0(t,0)$, with the 
initial condition
\be
C_{\vec k}(0) = \frac{T_I}{k^2 + \Lambda^2}.
\label{Incond.1}
\ee
Notice that $\omega_0(t)$ decreases 
from the initial positive value, vanishes at the characteristic time $t^*$, defined by
\be
S(t^*) = -r/g
\label{FRRMG.1}
\ee
and, after reaching a minimum, eventually vanishes again with a negative power law tail $t^{-1}$,
which can be derived analytically~\cite{CA,Bray} and is independent of the initial condition.

\begin{figure}[h]

	\vspace{1cm}

    \centering
   \rotatebox{0}{\resizebox{.4\textwidth}{!}{\includegraphics{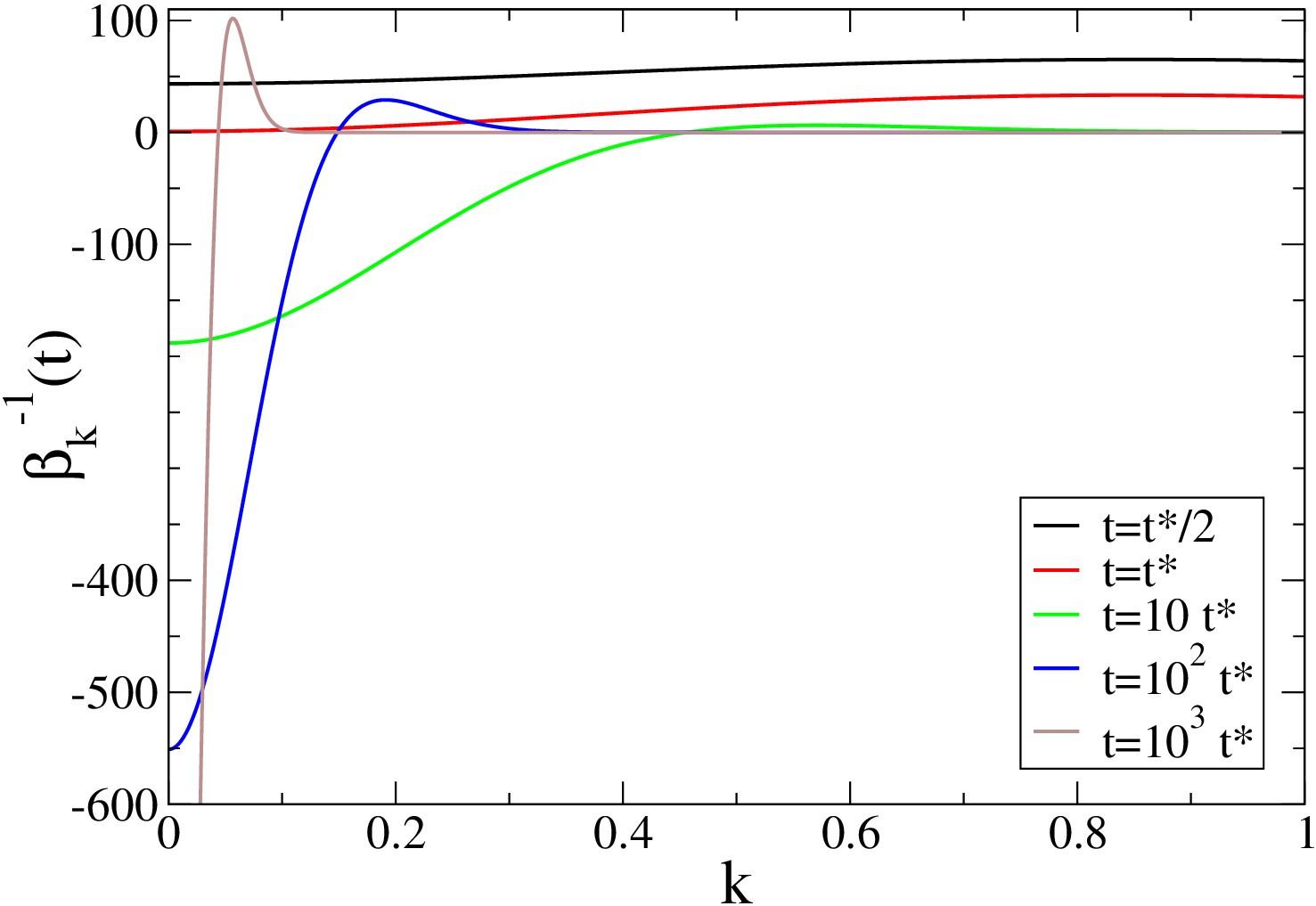}}}
\hspace{1cm}
   \rotatebox{0}{\resizebox{.4\textwidth}{!}{\includegraphics{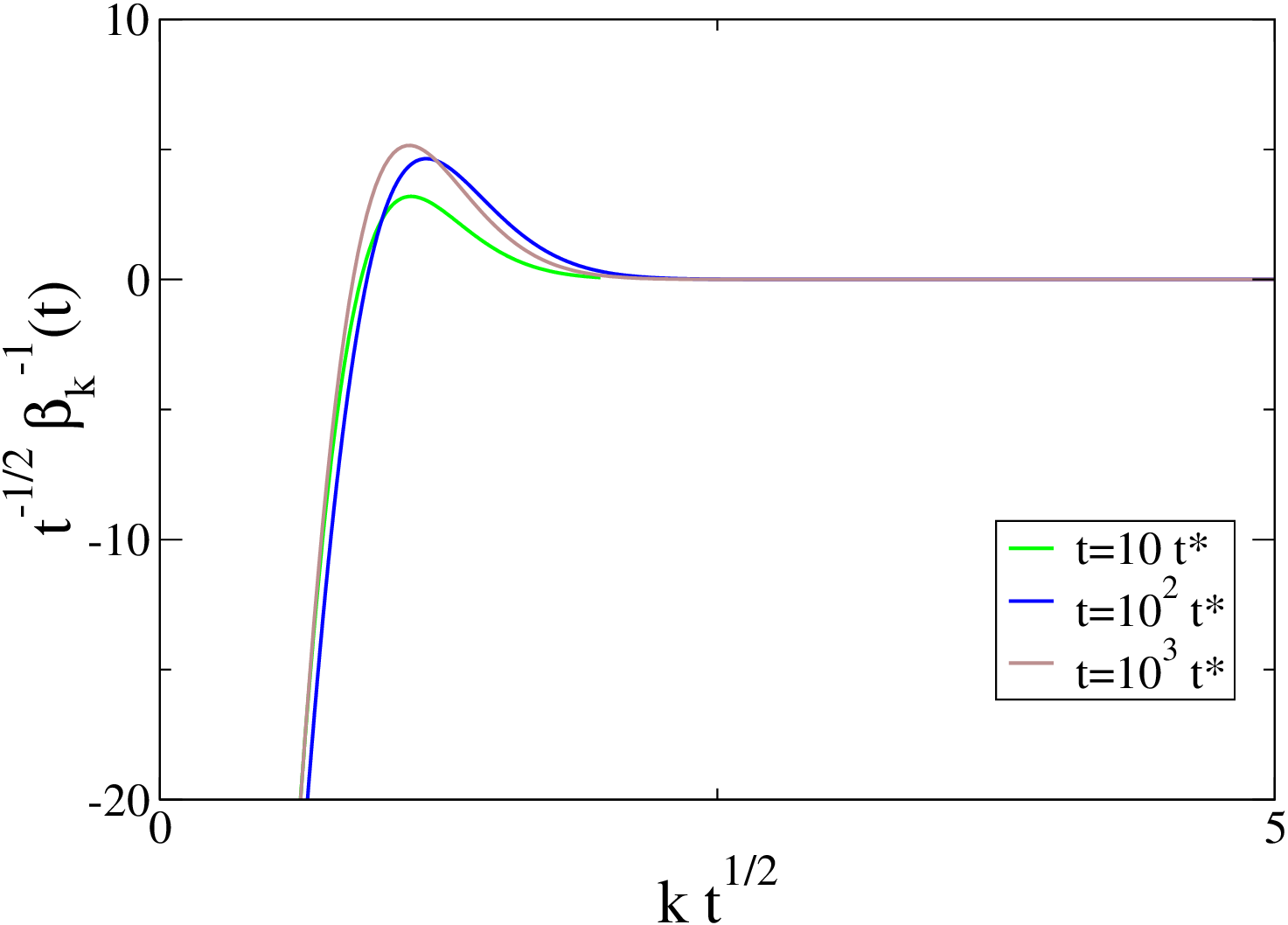}}}
\caption{Left panel: time evolution of the
spectrum of effective temperatures in the large $N$ model, quenched to $T_F=0$. 
Right panel: rescaled spectrum as in Eq.~(\ref{sc.3}). The curve for $t=10^3t^*$ is
indistinguishable from the plot of $g(x)$ defined by Eq.~(\ref{sc.4}).} 
\label{efftempninfty}
\end{figure}

Therefore, $\omega_0(t)$ remains negative from the time $t^*$ onward.
This means that for $t > t^*$ there appears a branch of unstable modes with
$\omega_k(t) < 0$, for $k < \sqrt{-\omega_0(t)}$. Recall that, having taken the
limit of $N$ large, we are looking at a single component under the action
of the mean-field interaction due to all the other components, which gives
rise to the time dependence of $\omega_0(t)$. As a consequence,
for $t > t^*$ a negative branch
appears in the temperature spectrum, as illustrated in the left panel of Fig.~\ref{efftempninfty}.
The evolution is as follows. Initially, at $t=0$, the spectrum is flat (not shown 
in the figure). Then, as soon as the system is quenched, equipartition
is broken, with a pattern showing the formation of a peak
which narrows and moves toward the origin. At early times, for $t < t^*$, all the modes cool with
the modes on the
sides of the peak cooling faster than the peak. However, for $t > t^*$ the peak reverses the trend by
growing and warming up, 
while the modes on its sides keep on cooling. Denoting by $k_H(t)$ the wave vector magnitude of the
{\it hot} peak, on the short wave length side $(k > k_H)$ 
equilibration to $T_F=0$ takes place. Instead, the long wave length modes $(k < k_H)$ keep on cooling indefinitely,
entering the negative branch for $t > t^*$. This is a remarkable feature, due to the nonlinearity of the
model, showing that energy is not just lost to the environment as in the Gaussian case, but that
in the process there is also reshuffling among the modes, 
while the global average energy relaxes to zero, as it will be clear shortly.

It is well known~\cite{Bray} that in the late stage of phase-ordering dynamical scaling
holds, in the sense that all lengths in the problem can be rescaled with respect
to the characteristic time dependent length $L(t) \sim t^{1/2}$, representing the typical size
of the growing domains of the ordered phases~\cite{domains}. The large $N$ model is one
of the few cases in which scaling can be derived analytically~\cite{CA,Bray}, yielding
for the equal time structure factor
\be
C_{\vec k}(t) = t^{d/2}f(kt^{1/2}),
\label{sc.1}
\ee
with $f(x) = (8 \pi)^{d/2}e^{-2x^2}$. Now, rewriting Eq.~(\ref{Gin.10}) as
\be
\dot{C}_{\vec k}(t) = -2\beta^{-1}_k(t),
\label{sc.2}
\ee
from Eq.~(\ref{sc.1}) follows immediately the scaling form of
the effective temperature 
\be
\beta^{-1}_k(t) = t^{d/2-1}g(kt^{1/2}),
\label{sc.3}
\ee
with
\be
g(x) = (8 \pi)^{d/2}(x^2-d/4)e^{-2x^2}.
\label{sc.4}
\ee
Thus, plotting $t^{1-d/2}\beta^{-1}_k(t)$ against $x=kt^{1/2}$, the curves for different values of $t$
should collapse on the master curve~(\ref{sc.4}). This is shown in the right panel
of Fig.~\ref{efftempninfty}, where the last curve, for $t=10^3t^*$, is undistinguishable
from the plot of the scaling function $g(x)$. 

Once scaling sets in,
the scaling function captures in one shot, so to speak, the entire asymptotic time evolution. The shape
of $g(x)$ tells us that the formation of the hot peak and, therefore, the breaking of equipartition are
permanent features of the phase-ordering process. This is in sharp contrast
with the behavior of the average energy, which, instead, equilibrates.
In fact, using Eq.~(\ref{Indv.2bis}), we have
\be
\langle {\cal H}_{\rm eff}[\mathbf{x},t] \rangle = \frac{\Upsilon_d}{2(2\pi)^d} \int_0^{\Lambda} dk \, k^{d-1}
\beta^{-1}_k(t)
\label{ggg.1}
\ee
and inserting the scaling form~(\ref{sc.3}) for the effective temperature there follows that
$\langle {\cal H}_{\rm eff}[\mathbf{x},t] \rangle$, for large $t$, vanishes like $t^{-1}$.
Thus, while the total average
energy shows a seemingly smooth relaxation to equilibrium,
the spectrum of effective temperatures reveals that the system is in fact stuck out of equilibrium,
as well illustrated by the pattern of breaking of equipartition.

\begin{figure}[h]

	\vspace{1cm}

    \centering
   \rotatebox{0}{\resizebox{.4\textwidth}{!}{\includegraphics{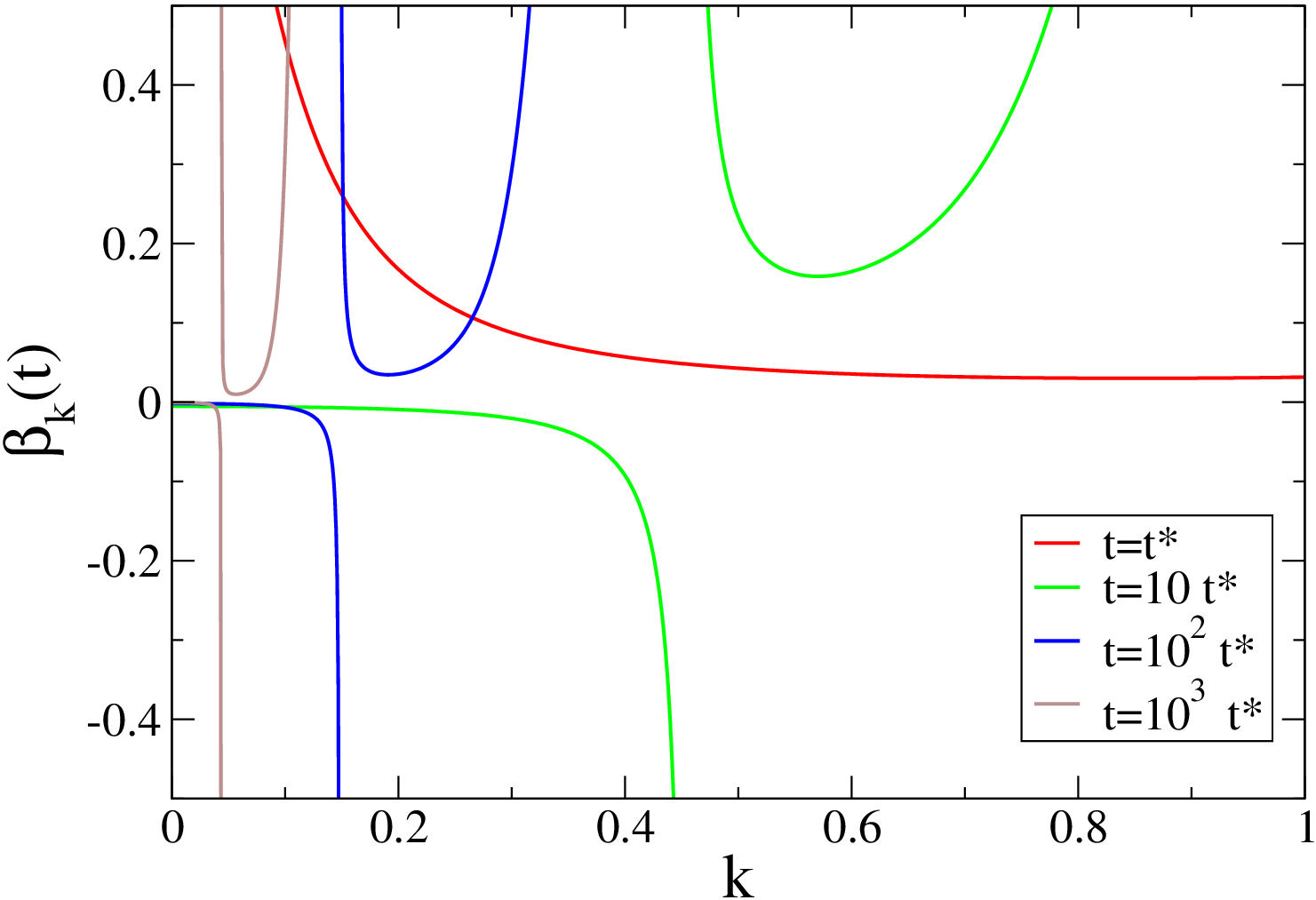}}}
\hspace{1cm}
   \rotatebox{0}{\resizebox{.4\textwidth}{!}{\includegraphics{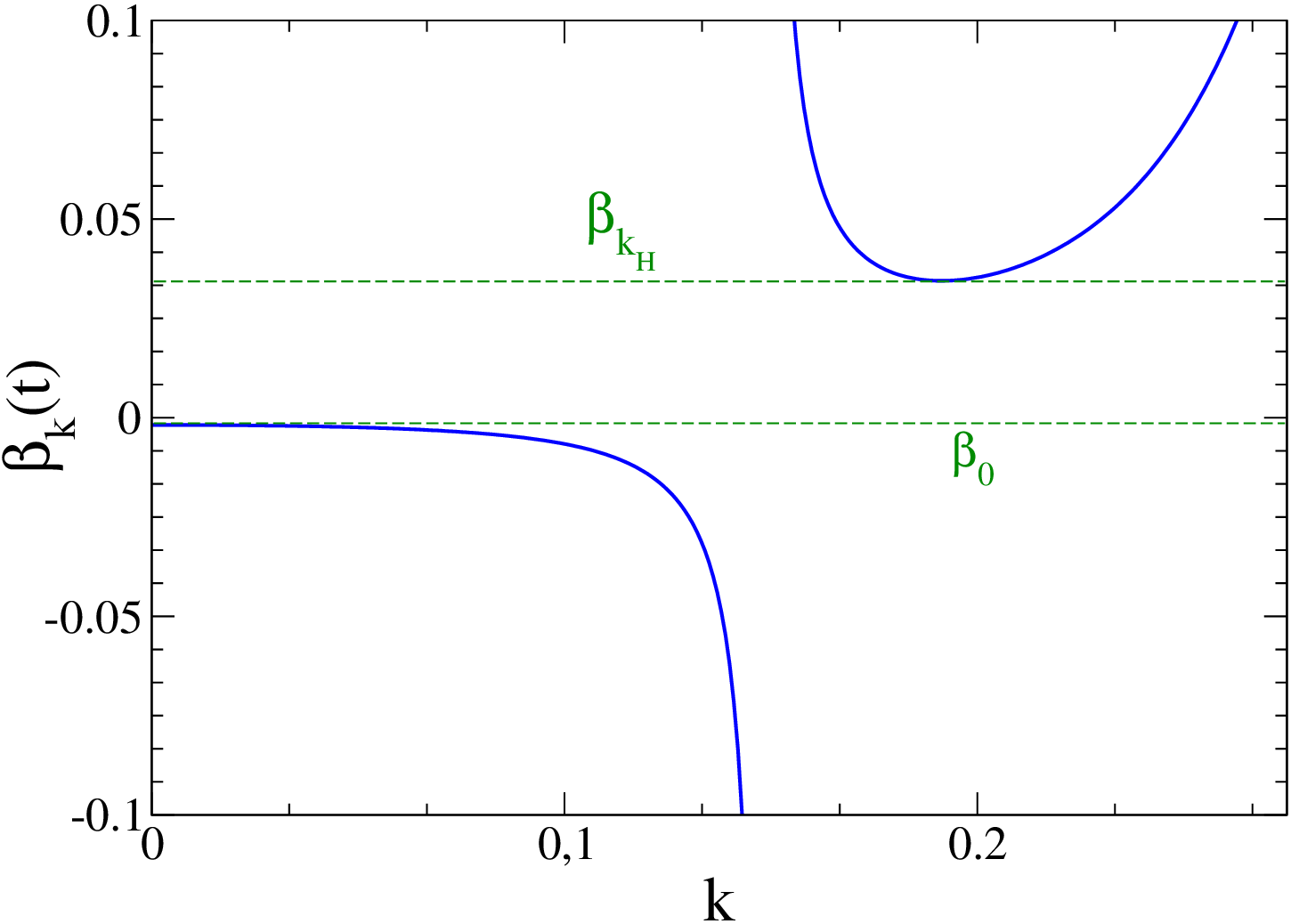}}}
\caption{Left panel: time evolution of the spectrum of inverse effective temperatures.
Right panel: magnification of the plot for $t=10^2t^*$. The edges of the positive and negative
branches are marked by $\beta_{k_H}$ and $\beta_0$.}
\label{invefftempninfty}
\end{figure}

\subsection{Energy fluctuations}

The difference between the two regimes, before and after $t^*$, is even more
evident in the spectrum of inverse effective temperatures 
(left panel of Fig.~\ref{invefftempninfty}).
Let us, then, look at the implications for the behavior
of energy fluctuations.

\vspace{5mm}

{\it $t < t^*$}

\vspace{5mm}

In this time regime the spectrum contains only the positive branch (left panel of Fig.~\ref{invefftempninfty}).
Hence, $F_{{\cal H}}(z,t)$ is defined for $z \leq \beta_{\rm min} = \beta_{k_H}$.
Proceeding as in 
Sec.~\ref{largevolume}, in place of Eq.~(\ref{condav.1}) we now have
\be
e = \frac{1}{V} \langle {\cal H}_{k_H} \rangle_{z} + F_{{\cal H}}(z,t), 
\label{FRRMG.3}
\ee 
where $F_{{\cal H}}(z,t)$ diverges
as $z$ approaches the lower edge of the spectrum at $\beta_{k_H}$, since $k_H(t) > 0$.
Thus, as long as $t < t^*$, the first term in the right hand side of the above 
equation is negligible and there is no condensation.

\vspace{5mm}

{\it $t > t^*$}

\vspace{5mm}

\begin{figure}
\begin{center}
\epsfysize=6.0cm \epsffile{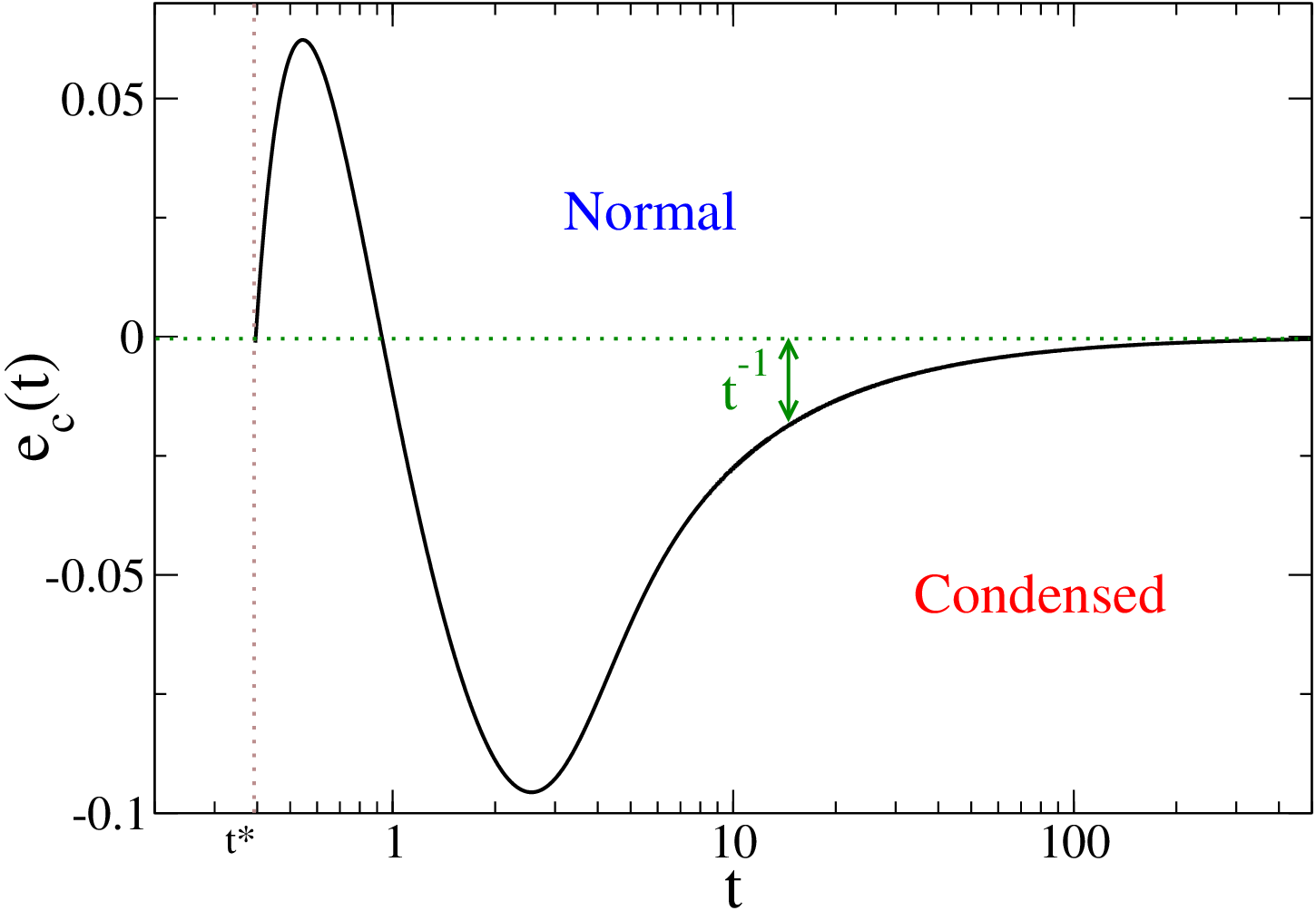} \caption{Energy phase diagram in the large $N$ model
quenched to $T_F=0$.} 
\label{enphasediagninfty}
\end{center}
\end{figure}

In this regime, as already mentioned and as shown in
Fig.~\ref{invefftempninfty}, the spectrum
exhibits both a negative and a positive branch, with edges at $\beta_{\rm max} = \beta_0$ and
$\beta_{\rm min} = \beta_{k_H}$.
Thus, the domain of definition of $F_{{\cal H}}(z,t)$ narrows to the finite interval $z \in (\beta_0,\beta_{k_H})$.
As $z \rightarrow \beta_0$, the lower bound of $F_{{\cal H}}(z,t)$ converges, yielding the critical threshold
\be
e_C(t) = F_{{\cal H}}(\beta_0,t).
\label{FRRMG.5}
\ee
The phase diagram, obtained by plotting $e_C(t)$ for $t > t^*$, is displayed in Fig.~\ref{enphasediagninfty}.
Comparing with the phase diagram of the Gaussian model, in the right panel of Fig.~\ref{ph_diag_S},
apart for the different shapes of the curves, the prominent qualitative difference is that the positions
of the normal and condensed phases are reversed with respect to the critical line. Namely, in the Gaussian 
case condensation takes place if an energy fluctuation is forced above a certain level, while the opposite
occurs in the large $N$ model. The origin of this radically different behavior is in the shapes of
effective temperatures spectra, with $\beta_0$ minimizing the positive branch
in the first case and maximizing the negative branch in the second one. Now, since the presence
of the negative branch is a consequence of the failure to equilibrate, the reversal of the phase
diagram is the most conspicuous manifestation that in the quench to below $T_C$ the system
does not equilibrate.

\section{Conclusions}
\label{conclusions}

We have investigated the fluctuations of energy and heat during the
relaxation following an instantaneous temperature quench.
The study has been carried out analytically in the Gaussian model, for a quench in the paramagnetic phase,
and in the large $N$ model, for a quench below $T_C$.
The main finding is the condensation of fluctuations in the
transient out-of-equilibrium regime, which occurs when the pattern of equipartition
breaking produces a spectrum with an extremal point at zero wave vector. Then,
condensation is driven by the same mathematical mechanism responsible of BEC.
The dynamical nature of the transition is well illustrated by the phase diagrams
extending in the time direction. 
We have treated the quenches in the paramagnetic and in the ferromagnetic phase, in order to study
the cases of a finite relaxation time and of an aging system, which remains permanently
out of equilibrium but not in a stationary state. In both cases, the concept of effective temperature
plays an essential role in the characterization of the distance from equilibrium and
in uncovering the mechanism of the transition.
 
At first sight, it may seem surprising that a transition occurs in a non interacting
system, such as the Gaussian model. The apparent puzzle is solved in the framework
of ensemble duality, showing that condensation of fluctuations
in the prior, as a large deviation manifestation, corresponds to condensation on
average in the biased ensemble, where
the imposition of the bias amounts to a mean-field interaction. 
Finally, let us remark that
the interplay of the different ensembles related 
by the rate function, is very interesting in itself, since ensembles are the byproduct of
the observation procedures. In this respect, much effort has been devoted to
use ensemble duality to find implementable ways of making accessible to
observation rare events.  However, at least in principle, the connection could work
also in the opposite way, in those cases where the imposition of the bias could
be more difficult to realize than the observation of large deviations.

Finally, the natural question is about the generality of the results 
reported in this paper. They ought to be generic for separable systems, which is a prerequisite 
for a meaningful definition of an effective temperature per normal mode. Though blurred, the picture 
ought to survive also in weakly interacting systems, much in the same way as for BEC of cold atoms.
More speculative is the question of the possible observation, since it involves recording of the
fluctuations, whose feasibility clearly depends on how far is the critical threshold from
the average observed behavior and, therefore, can be addressed only on a case-by-case basis.

\vspace{2cm}

\noindent {\bf e-mail addresses} - mrc.zannetti@gmail.com, corberi@sa.infn.it, 

\noindent gonnella@ba.infn.it, antps@hotmail.it


\begin{thebibliography}{99}
\bibitem{Berlin} The mechanism for the ``sticking'' of the saddle point solution is analysed in
the solution of the spherical model in Ref.~\cite{BK}.
\bibitem{domains}
Strictly speaking, no domains are formed in the large $N$ model (see Ref.~\cite{castellano}).
In this model it is more correct to address $L(t)$ as the time dependent correlation length.
\end{thebibliography}
\end{document}